# Renewable levelized cost of energy available for export: An indicator for exploring global renewable energy trade potential

Xiaoming Kan[a,*] Lina Reichenberg[a] Fredrik Hedenus[a] David Daniels[a]

[a]Department of Space, Earth and Environment, Chalmers University of Technology, 41296 Gothenburg, Sweden

## ABSTRACT

Renewable energy resources are widely available, yet they are unevenly distributed globally. In a renewable future, countries lacking high-quality renewable resources may choose to import energy from other countries. To assess the resource-dependent and techno-economic basis for global renewable energy trade and identify potential importers and exporters, this study introduces two new metrics: Renewable Levelized Cost of Energy available for Export ($RLCOE_{Ex}$) and Potential Energy Export Volume ($PEEV$). These metrics are computed based on regional resource potential, domestic energy demand and varying financial costs across countries, without the need for any energy system modeling. By applying these two metrics to 165 countries/regions, we identify countries with significant potential for exporting renewable energy (e.g., the US, China) and those that lack the domestic resources to satisfy demand (e.g., South Korea, Japan). The $RLCOE_{Ex}$ and $PEEV$ metrics are validated through a separate analysis, employing a comprehensive energy system model for each country/region.

## 1. Introduction

Limiting the increase in global average temperature to “well below” 2°C entails an energy transition towards nearly zero or even negative $CO_2$ emissions by mid-century [1, 2]. Electrification of transportation, heating and industrial sectors, directly or via electricity-derived fuels, will require a substantial increase in electricity supply [3, 4]. Following sustained cost reductions and rapid diffusion into the power generation mix, renewable energy technologies such as wind power and solar photovoltaic (PV) may serve as the cornerstone for the future low-carbon electricity system [3, 5-7]. Though renewable resources are broadly available, they are unevenly distributed globally [8, 9]. Therefore, similar to other natural resources, some countries have more potential than others to meet their energy demands using domestic resources. In a decarbonized future, countries that lack enough high-quality renewable resources to meet their domestic demand can either invest in nuclear power, deploy carbon capture and storage (CCS) with fossil fuel power plants or import energy from other countries. Decarbonizing some end-use sectors may also increase demand for hydrogen and other synthetic fuels produced using renewable electricity as a “feedstock” [3, 10]. Thus, countries whose renewable energy production exceeds domestic demand could become energy supply nodes both for electricity and for electricity-derived fuels [11, 12]. In this study, we evaluate the global potential of renewable energy and identify countries that could serve as potential importers or exporters. To accomplish this, we introduce two novel metrics: **Renewable Levelized Cost of Energy available for Export** ($RLCOE_{Ex}$) and **Potential Energy Export Volume** (*PEEV*). Our analysis considers regional resource endowments, land availability for wind and solar power installations, domestic energy demand and country-specific discount rates, which reflect the heterogeneous financial costs across countries [13]. As climate mitigation scenarios anticipate significant expansion in wind and solar energy [14], and our primary objective is to elucidate concepts rather than predict the eventual winners among technologies, this study primarily focuses on wind and solar energy, in addition to existing hydropower, for the sake of simplicity.

* Correspondence: kanx@chalmers.se

Numerous studies have explored wind and solar energy potential [8, 9, 15-24], with some delving into global aspects beyond pure technical potential. While these studies [8, 9, 15-24] focused on the renewable resources themselves, other studies complemented the physical and economic potential analyses with other factors to identify potential importers and exporters of hydrogen. Notably, Pflugmann and De Blasio [25] incorporated factors such as domestic energy demand, freshwater availability for hydrogen production and infrastructure capabilities into their analysis, and proposed that the US and Australia have the potential to become leading hydrogen exporters. In addition to the factors considered in Ref. [25], Tonelli et al. [26] considered the water footprint of wind, solar and electrolyzer infrastructures in their analysis, and identified Southern Africa, South America, Canada and Australia as promising leaders in hydrogen export. The IRENA report [27] took a further step by accounting for heterogeneous financial costs when estimating hydrogen supply costs and potential. Based on a survey with experts in the field, Hjeij et al. [28] developed a hydrogen export competitiveness indicator that integrated resource availability and potential, economic and financial potential, political and regulatory status and industrial knowledge. Their analysis identified the US, Australia, Canada and China as the most competitive countries in hydrogen export. Other studies employed complex energy system models to estimate regional hydrogen supply curves for Europe [29] and the Middle East and North Africa [30]. Additionally, some other studies [31-34] applied a global energy system optimization model to investigate potential electricity or hydrogen trade between countries [31-34].

Unlike previous studies that primarily focused on the heterogeneity of the renewable energy resource itself [8, 9, 15-24], this study also recognizes national differences with respect to the financial ability to develop the resources, as well as the domestic energy demand. Moreover, we do not narrow the analysis to focus solely on the vagaries of trade in, e.g., hydrogen, like in Refs. [34, 35]. Thus, this study occupies the space between studies of renewable potential [8, 9, 15-24] and those using complex energy system optimization models to map out the specifics of trade [31-34]. While intuitive economic comparison metrics like Levelized Cost of Energy (*LCOE*) are insufficient predictors of future global energy system growth, the complexity of global energy system optimization models can also create barriers to understanding: the optimization process typically selects only one or a few trade routes/partnerships, while in reality, there are numerous possibilities. By contrast, our approach offers a global techno-economic backdrop against which further socio-political analysis can be conducted to identify potential trade partners. We achieve this by incorporating the diverse resource endowments (solar insolation and wind speed) and socio-economic realities (financial costs and energy demands) across the world into two metrics to assess the resource quality and quantity for trade. Our metrics are relatively simple compared to detailed energy system models [29, 30, 34] and the more complex indicators introduced by Refs. [25-28]. Still, they capture the factors that Hjeij et al. [28] proposed as the most influential. It is important to note that the $RLCOE_{Ex}$ and $PEEV$ metrics are agnostic to the specific form of energy trade. Therefore, they are more generic indicators for assessing energy trade potential than the hydrogen export competitiveness indicators developed in Refs. [25-28].

This study offers three key contributions. First, from a methodological perspective, we introduce two new metrics: $RLCOE_{Ex}$ and $PEEV$. These metrics serve as comprehensive indicators for the potential future trade of renewable energy, and they are easy to compute since they do not require any energy system modeling. Second, by applying the $RLCOE_{Ex}$ and $PEEV$ metrics to 165 countries/regions in the world, we provide a global view of the relative competitiveness of national renewable energy exports between countries. Third, compared with previous hydrogen export competitiveness indicators [25-28], we validate the utility of the $RLCOE_{Ex}$ and $PEEV$ metrics by comparing them with the marginal hydrogen cost and potential hydrogen export volume for each country, calculated using a comprehensive energy system model.

## 2. Results

We start by outlining the method for calculating the $RLCOE_{Ex}$ and $PEEV$ metrics, and we present the estimates of their values for different countries/regions in the world. Next, we illustrate the influence of diverse financial costs on both $RLCOE_{Ex}$ and $PEEV$. Finally, we validate the utility of these two metrics by comparing them with the results obtained from country-specific, techno-economic energy system modeling analyses.

### 2.1 Two simple metrics to measure renewable export potential: $RLCOE_{Ex}$ and $PEEV$

$RLCOE_{Ex}$ represents the cost of providing an additional unit of energy for export after annual domestic energy demand is met. We first calculate the $LCOE$ of renewable energy for every grid cell ($0.01° \times 0.01°$) in each country. Subsequently, we arrange the $LCOE$ in ascending order to generate the national renewable energy supply curve (Figure 1). Then, we identify the annual domestic energy demand (both electricity and hydrogen[1] demands), indicated by the red line at 1.0 on the x-axis in Figure 1. The intersection point of the annual supply and annual demand on the supply curve represents the $\boldsymbol{RLCOE_{Ex}}$. Hence, $RLCOE_{Ex}$ measures the cost at which surplus energy beyond domestic demand may be exported, as well as the marginal cost for a country to supply its entire energy demand using only domestic solar, wind and hydro resources (see Methods 4.3 for further details). The $RLCOE_{Ex}$ metric takes into account, in addition to wind and solar conditions and country-specific discount rates, demographic factors (energy demand and available land for wind and solar power installations) and the contribution from existing hydropower. Similar to the regular $LCOE$, $RLCOE_{Ex}$ is not the actual energy cost in the energy market. Instead, it reflects the investment and operational costs linked to the energy generation technology required to produce an additional unit of energy for export once the annual domestic demand has been met. Therefore, it should not be interpreted as the marginal cost to produce electricity or hydrogen, which encompasses additional costs (see Section 2.3 and Discussion). By setting a threshold on the national renewable energy supply curve, it is possible to estimate the total renewable energy production ($Pt$) under the threshold (Figure 1). Subtracting the domestic energy demand from $Pt$ gives us the $\boldsymbol{PEEV}$. Given the uncertainties in estimates of the evolution of future global renewable energy costs, the absolute values of $RLCOE_{Ex}$ and $PEEV$ are less important than the relative order of countries based on their $RLCOE_{Ex}$ and $PEEV$ values.

[1] Hydrogen may be used both directly (e.g., in industry) or as a feedstock to produce synthetic fuels.

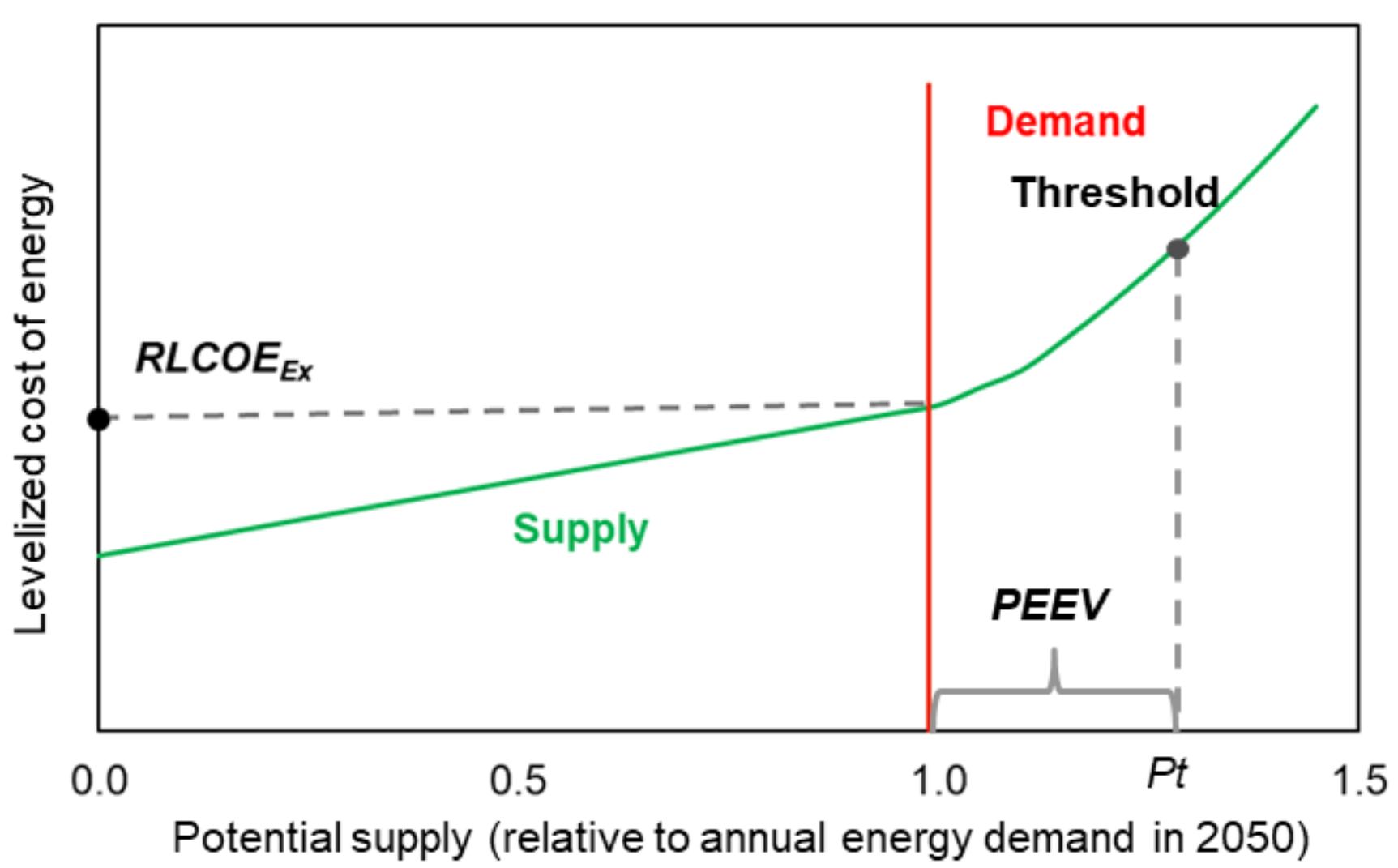

**Figure 1** A schematic diagram for estimating $RLCOE_{Ex}$ and $PEEV$. The x-axis represents the renewable energy supply potential relative to the estimated annual energy demand in 2050.

We estimate $RLCOE_{Ex}$ at the country level (small and medium-sized countries) or, for some big countries, subnational level, see Figure 2. Saudi Arabia, Chile, Morocco and the majority of the US, China, Mexico, Brazil and Australia show relatively low $RLCOE_{Ex}$ values, thus being potential exporters of renewable energy. The export possibilities are especially favorable for China, where neighboring countries display high $RLCOE_{Ex}$, or, in the case of Japan and South Korea, are unable to meet their demands using domestic resources. Due to unfavorably high financial costs, some African and Latin American countries exhibit relatively high $RLCOE_{Ex}$ values. These values are on par with those of Central or even Northern European countries, despite the latter having significantly less favorable solar conditions.

Africa is sometimes cited as a continent that may rely on distributed rather than centralized power, since the solar resource is abundant and evenly distributed [36]. However, Figure 2 shows considerable heterogeneity within the continent, with several North African countries displaying a low $RLCOE_{Ex}$, while the costs for some countries in the central part are notably high. Such uneven $RLCOE_{Ex}$ values may provide incentives for developing long-distance power transmission grids for electricity trade. The heterogeneity in $RLCOE_{Ex}$ is also observed within large countries such as the US and China. For many countries, the $RLCOE_{Ex}$ is below 20 \$/MWh, with the lowest cost reaching 13 \$/MWh (Figure 2). In stark contrast, around 30% of the countries are not self-sufficient or have a $RLCOE_{Ex}$ greater than 30 \$/MWh. The large number of countries with comparably low $RLCOE_{Ex}$ may provide plenty of energy trade options for countries with insufficient renewable resources.

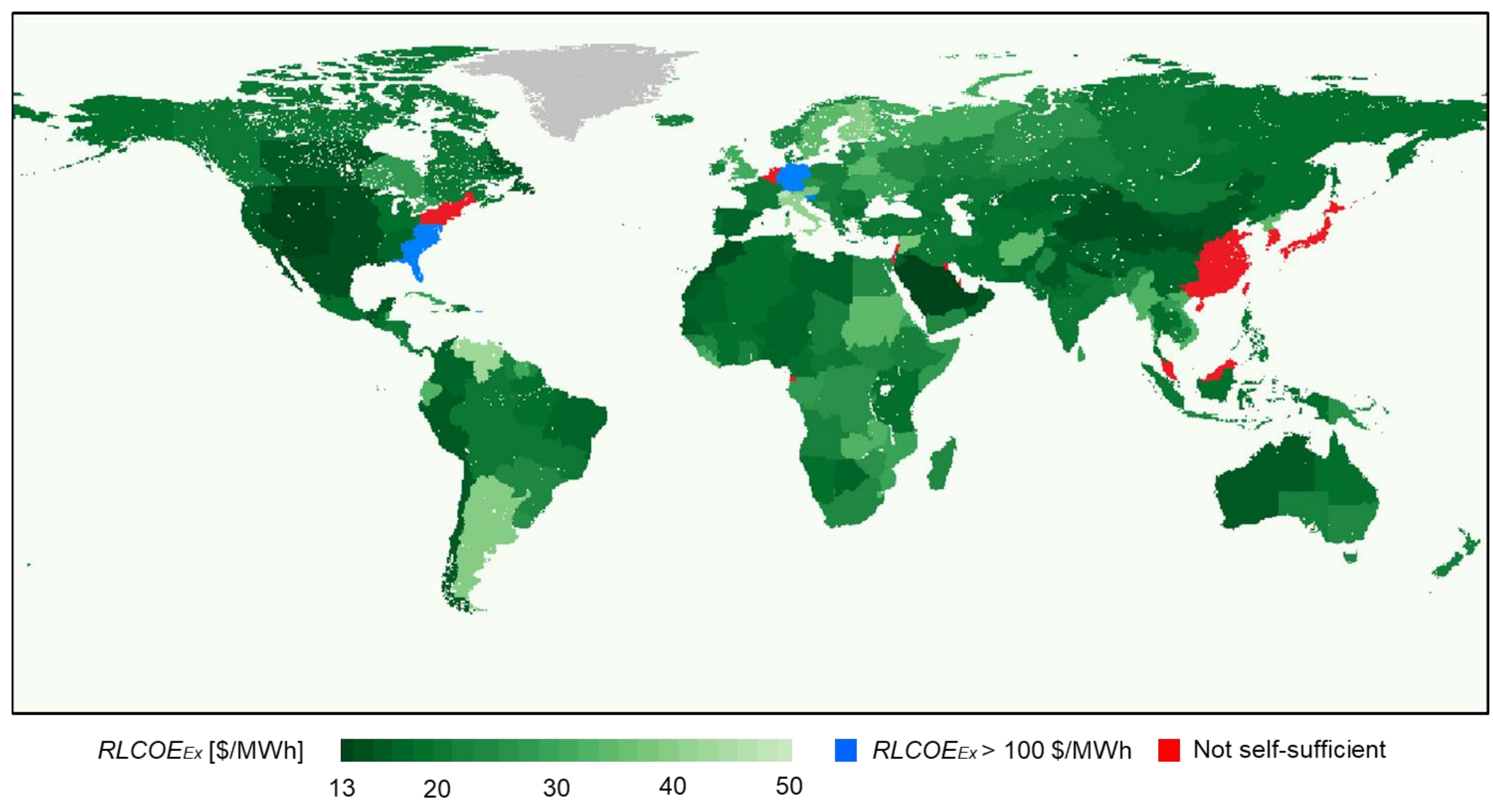

**Figure 2** $RLCOE_{Ex}$ for 2050. $RLCOE_{Ex}$ is estimated for most countries/regions in the world based on projected energy demand in 2050 and using current country-specific discount rates.

Apart from assessing the relative cost-competitiveness of exporting renewable energy, we also analyze the *PEEV* (in PWh) following the fulfillment of domestic energy needs for each country. This analysis assumes a threshold on the national renewable energy supply curves, under which there is a demand from other countries to import energy (see Figure 1). Under a threshold of 35 \$/MWh, the countries that exhibit the highest *PEEV* are the US, China, Brazil and Saudi Arabia (Figure 3). It is noteworthy that either the US or China alone exhibits a *PEEV* comparable to the current global electricity consumption (25 PWh) [37]. Under a lower threshold of \$25/MWh, the US, China, and Saudi Arabia dominate the potential exporting countries, with a potential export volume accounting for 30% of the total global export potential (Figure S1). These countries possess favorable wind and solar resources and display low financial costs for investments. This advantageous combination allows them to produce a significant amount of low-cost renewable energy that surpasses their domestic demand. Unsurprisingly, the $PEEV$ increases as the value of the threshold on the supply curve rises (Figure S1).

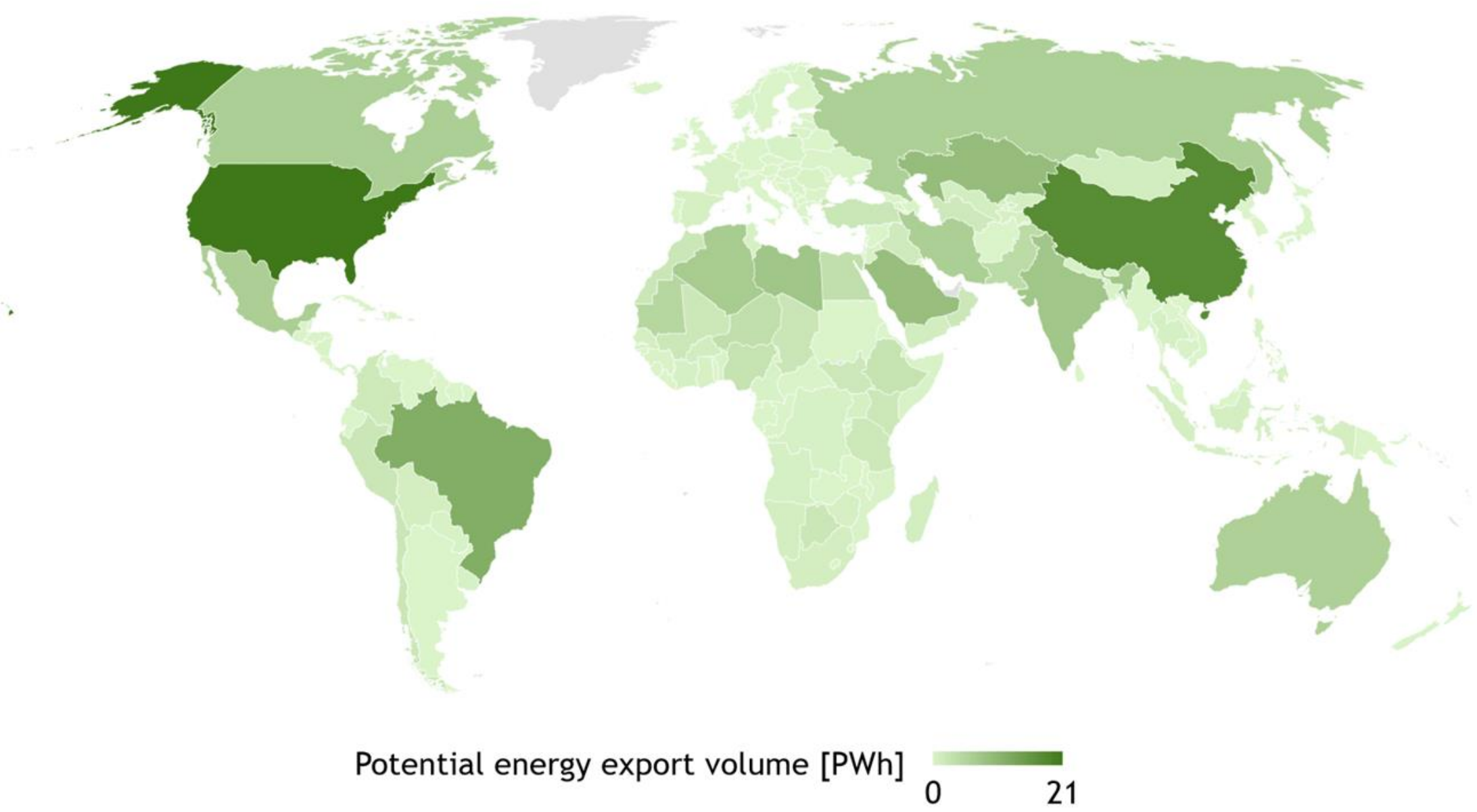

**Figure 3** The potential energy export volume under a threshold of 35 $/MWh on the national renewable energy supply curve

## 2.2 Validation of $RLCOE_{Ex}$ and $PEEV$: the cost to produce hydrogen and the potential hydrogen export volume

The $LCOE$ of renewable energy, on which the $RLCOE_{Ex}$ and $PEEV$ metrics are based, does not consider system-related costs for variation management [39, 40], nor does it factor in the expenses associated with producing electricity-derived fuels, such as the investment cost of electrolyzers. Therefore, we validate the $RLCOE_{Ex}$ and $PEEV$ results by comparing them with marginal hydrogen cost and potential hydrogen export volume for each country calculated by a detailed energy system model [41] (see Methods 4.4). The marginal hydrogen cost represents the cost of producing an additional unit of hydrogen for export after the domestic hydrogen demand has been met. Unlike the simple $RLCOE_{Ex}$ metric, the marginal hydrogen cost represents the cost for hydrogen at the exporting node. Figure 4 shows the relationship between the marginal hydrogen cost and $RLCOE_{Ex}$ for most countries in the world. The marginal hydrogen cost exhibits a high degree of correlation with $RLCOE_{Ex}$, as indicated by a Spearman correlation coefficient of 0.72 (Figure 4). Furthermore, the relationship between $RLCOE_{Ex}$ and marginal hydrogen cost exhibits a similar pattern across countries, with the marginal hydrogen cost being approximately twice that of $RLCOE_{Ex}$ (Figure 4). This relationship aligns with the findings of other studies regarding the proportion of electricity generation cost in the total hydrogen cost [42-44]. Therefore, the relative cost-competitiveness of hydrogen export among countries based on $RLCOE_{Ex}$ analysis remains valid even when factoring in system integration costs and the expenses associated with producing a specific quantity of hydrogen. $RLCOE_{Ex}$ can indeed serve as a valuable indicator for comparing energy export costs across different countries and identifying potential importers and exporters.

The main reason that the correlation coefficient is not higher than 0.72 is that the system integration costs vary across countries depending on the availability of domestic resources [45]. The presence of hydropower and the geographical size for spatial smoothing can influence the system integration costs and, consequently, the actual energy costs [7, 39]. If a country has abundant hydropower, the system integration cost is likely to be relatively low due to the inherent flexibility and storage capability provided by hydropower [46]. For example, the two red dots in Figure 4 represent Bhutan and Laos, where the marginal hydrogen cost is lower than 20 $/MWh thanks to the abundant hydropower resources (that exceed domestic electricity demand) in these two countries.

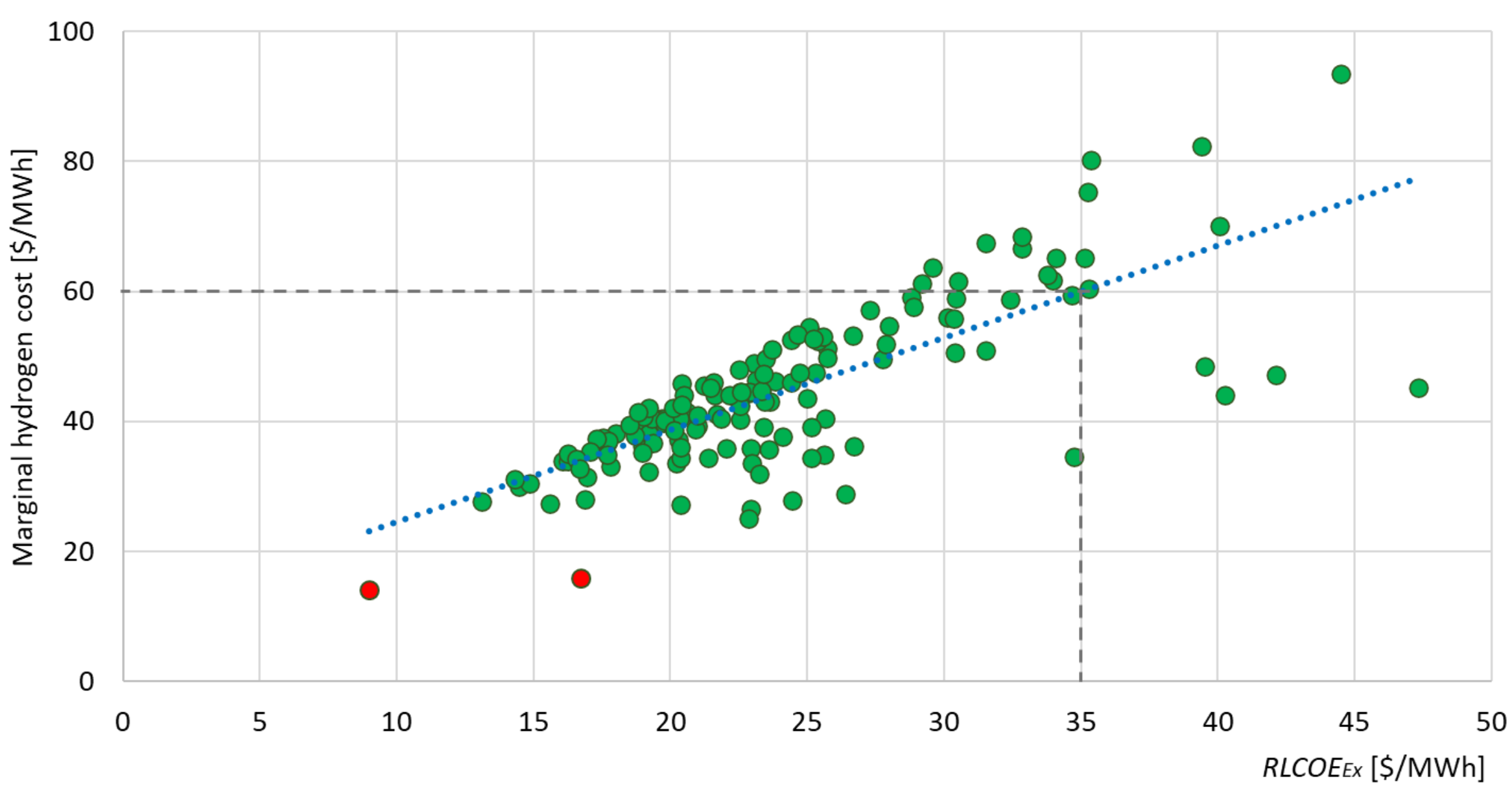

**Figure 4** Relationships between the marginal hydrogen cost and $RLCOE_{Ex}$. To enhance visualization clarity, countries that are not self-sufficient or exhibit exceptionally high costs due to land constraints are excluded. Only countries with a $RLCOE_{Ex}$ below 50 $/MWh are considered. The calculation of *PEEV* and the potential hydrogen export volume is carried out for countries located within the rectangle marked by the grey dashed lines.

We also validate *PEEV* by comparing it with the potential hydrogen export volume estimated with a detailed energy system model (see Methods 4.4). Figure 5 illustrates the relationship between the potential hydrogen export volume and *PEEV*. We evaluate the hydrogen export volume per country for a marginal hydrogen cost below 60 $/MWh and the *PEEV* for a $RLCOE_{Ex}$ below 35 $/MWh, see the marked rectangle area in Figure 4[2]. The threshold value for the marginal hydrogen cost was chosen based on the trendline in Figure 4, where an $RLCOE_{Ex}$ value of 35 \$/MWh corresponds to a marginal hydrogen cost of 60 \$/MWh. The potential hydrogen export volume shows a positive correlation with $PEEV$, with a remarkably high Spearman correlation coefficient of 0.97 (Figure 5). This finding further emphasizes that $RLCOE_{Ex}$ and $PPEV$ together are valuable, not only for comparing energy export costs but also for assessing the volumes of export across different countries. Note that the $RLCOE_{Ex}$ and $PEEV$ metrics are agnostic to the form of energy trade. Our modeling analysis of hydrogen serves as an example to validate their effectiveness.

[2] This means that countries that display a $RLCOE_{Ex}$ value above 35 $/MWh or a marginal hydrogen cost above 60 $/MWh are not considered for the validation of *PEEV*.

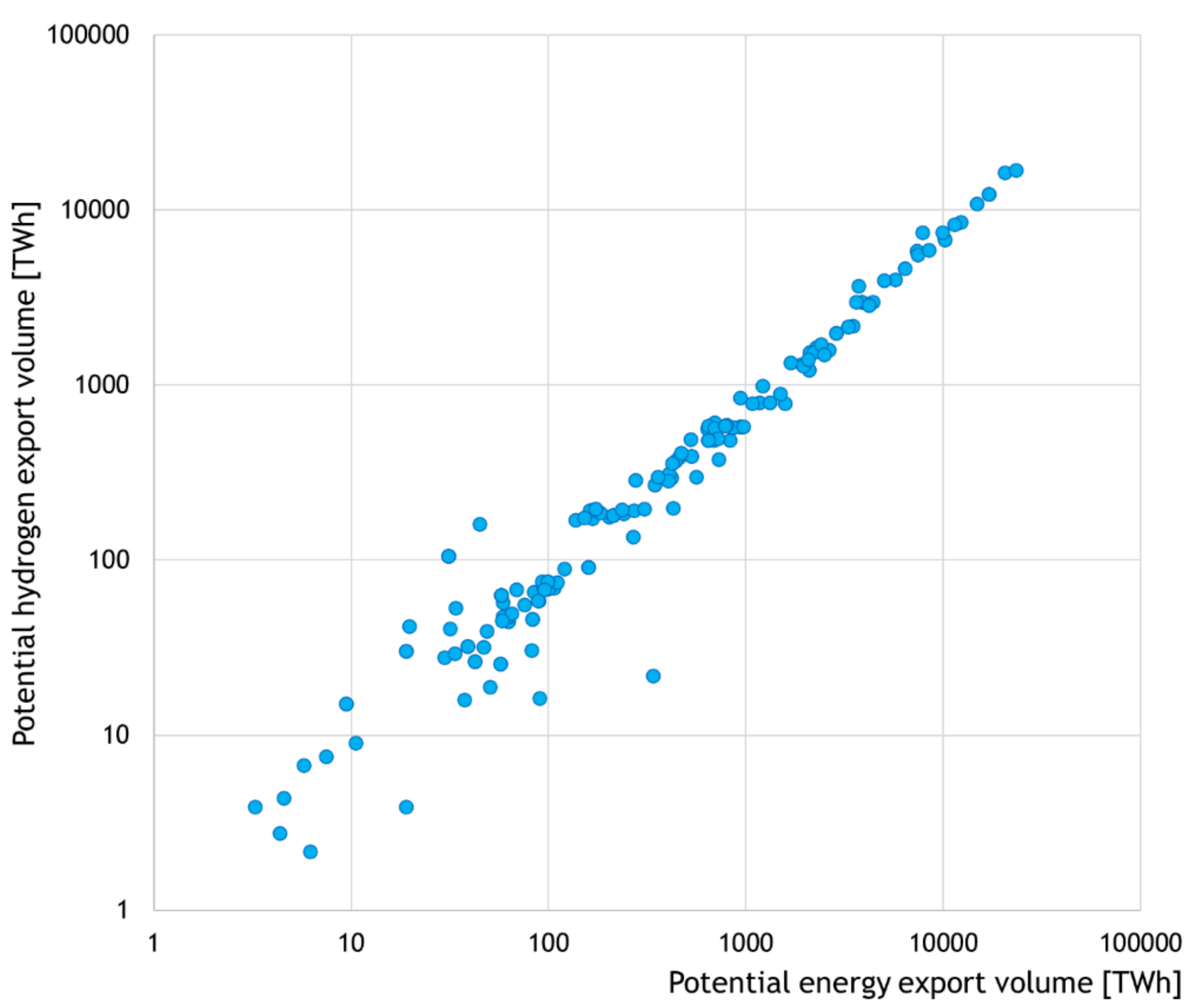

**Figure 5** The relationship between potential hydrogen export volume and $PEEV$ for the 165 countries/regions included in this analysis. Both the x-axis and the y-axis are displayed using a logarithmic scale.

## 2.3 How much do financial costs matter? - Impact of heterogeneous discount rates on renewable export potential

To investigate the impact of heterogeneity in financial costs on the quality and quantity of renewable energy trade, we compare $RLCOE_{Ex}$ and $PEEV$ using both country-specific and uniform discount rates. The country-specific discount rates are obtained by adding the country-specific risk premiums [38] to a "risk-free" baseline discount rate (5%). As illustrated in Figure 6a, in an optimistic future where all countries are harmonized to the same discount rate (5%), the $RLCOE_{Ex}$ for countries with the highest discount rates today decreases significantly (by more than 60%) compared to the values calculated with country-specific discount rates. Notably, the $RLCOE_{Ex}$ for Sudan and Venezuela is more than halved, indicating a substantial reduction in the cost of exporting renewable energy if these countries were to experience improved socio-political conditions and reduced financial costs. In addition to lowering the $RLCOE_{Ex}$, a lower discount rate also results in an increase in $PEEV$. This is because $PEEV$ is the surplus energy exceeding domestic demand and having an $LCOE$ below 35 $/MWh. This effect is particularly significant for countries with large geographical areas, such as Mongolia and Chad, where the $PEEV$ more than doubles (Figure 6b). These countries possess abundant high-quality renewable resources, but their renewable energy development may be hindered by high financial costs. Overall, we see that the assumption on discount rate significantly influences the assessment of both the quality and quantity of renewable energy potential. For potential implications, please refer to the Discussion section.

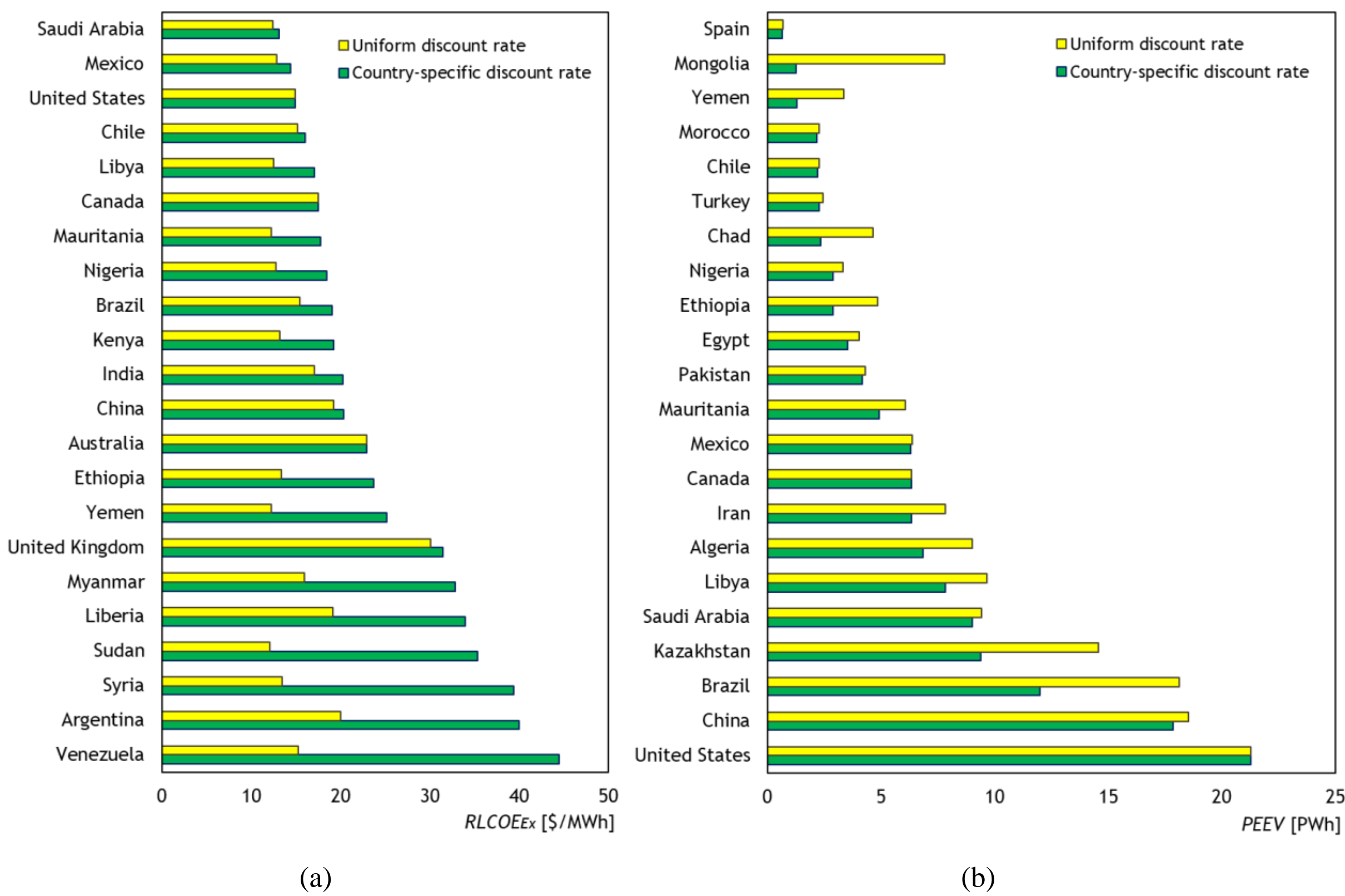

**Figure 6** $RLCOE_{Ex}$ (a) and $PEEV$ under a threshold of 35 $/MWh (b) estimated with country-specific (green) and uniform (yellow) discount rates for some selected countries.

## 2.4 Sensitivity analysis: double electricity and hydrogen demands

To account for potential large-scale electrification and the production of electricity-derived fuels for energy sectors beyond the electricity system, we conduct sensitivity analyses by doubling the electricity and hydrogen demands, one at a time. Doubling the electricity demand does not affect the $RLCOE_{Ex}$ significantly for most of the countries, except for countries with a high population density. Some of these countries are no longer self-sufficient, see Figure S2 & S3. In particular, a substantial impact is observed in European countries due to the change in electricity demand assumptions. The United Kingdom, Italy, Austria and the Czech Republic exhibit $RLCOE_{Ex}$ values exceeding 200 $/MWh, while

Germany, Switzerland and Slovenia are no longer self-sufficient with domestic renewable resources (Figure S3). As for hydrogen, if the demand is doubled, the marginal hydrogen cost correlates better with $RLCOE_{Ex}$, with a Spearman correlation coefficient reaching 0.9 (Figure S4), compared to the correlation coefficient of 0.72 with the original assumption on hydrogen demand (Figure 4). This suggests that an increased domestic hydrogen demand does not diminish the validity of $RLCOE_{Ex}$ as a metric for assessing the cost-competitiveness of hydrogen exports between countries.

## 3. Discussion

This study introduces two new metrics, $RLCOE_{Ex}$ and $PEEV$, to comprehensively evaluate the quality and quantity of renewable energy exports worldwide. These metrics take into account regional renewable energy resource (wind, solar and hydro) endowments, country-specific discount rates, land availability for wind and solar power installations and domestic energy demand. These metrics are easy to compute using openly available data, without the need for complicated energy system models. By applying the $RLCOE_{Ex}$ and $PEEV$ metrics to most countries in the world, we offer a comprehensive global perspective on the relative competitiveness of national renewable energy exports across countries.

The $RLCOE_{Ex}$ and $PEEV$ metrics extend beyond the mere physical potential studies [17-19, 24] as they assess the availability of energy for export after fulfilling domestic energy requirements, regardless of the potential energy trade form. These two metrics offer a broader scope than other comprehensive indices developed by Hjeij et al. [28] and the IRENA report [27] for evaluating the competitiveness of hydrogen exports. Like the indices in Hjeij et al. [28] and the IRENA report [27], they provide a simple and transparent approach for deriving insights about the potential for global renewable energy trade, and they do not require a full-scale global energy system modeling analysis like in Refs. [31-34]. Unlike Hjeij et al. [28], the IRENA report [27] and Tonelli et al. [26], we validate the utility of our metrics with a comprehensive energy system modeling analysis of all 165 countries/regions included in our study.

The $RLCOE_{Ex}$ and $PEEV$ metrics focus solely on the costs associated with power generation assets. A large part of the cost to satisfy demand in a system based on variable renewables consists of the cost to provide the so-called system integration [39, 40], namely variation management, including storage and backup capacity to address the intermittency of wind and solar power production [6, 7, 47]. Additionally, the decarbonization of certain end-use sectors is likely to drive an increased demand for electricity-derived fuels like hydrogen. To investigate whether the inclusion of additional system integration costs for wind and solar power, as well as hydrogen production, could influence the conclusions regarding potential importers and exporters based on $RLCOE_{Ex}$ and $PEEV$ analysis, we use a detailed techno-economic cost optimization model for capacity investment and dispatch [41] to model each country covered in our study (see Methods 4.4). We find that the $RLCOE_{Ex}$ is indeed highly correlated with the marginal cost to produce hydrogen at future predictions for hydrogen demand (correlation coefficient 0.72, Figure 4). The marginal cost of hydrogen is approximately double that of $RLCOE_{Ex}$ across the countries (Figures 4&S4). Additionally, we observe a strong correlation between the $PEEV$ and the potential hydrogen export volume (correlation coefficient of 0.97, Figure 5). This reinforces the value of $RLCOE_{Ex}$ and $PEEV$ as effective tools for assessing the competitiveness of renewable energy exports across different countries.

In contrast to the comprehensive hydrogen export competitiveness index developed by Hjeij et al. [28], which considers resource availability and potential, economic and financial potential, political and regulatory status and industrial knowledge, this study places a greater emphasis on renewable resource potential and economic factors of each country. This choice hinges partly on the perspective regarding future hydrogen or other electricity-derived fuels as predicated on, but by no means determined only by, the physical resource endowment [48]. This perspective aligns partially with the methodology employed in Hjeij et al. [28], where they assigned importance to factors related to resource availability and financial stability, accounting for roughly 75% of their index. The $RLCOE_{Ex}$ and $PEEV$ metrics do not incorporate certain factors such as freshwater scarcity, as emphasized in Ref. [25, 26], which can be addressed through alternatives like desalination at a relatively cheap cost compared to the total cost of hydrogen production for regions not significantly far from the coast [49]. However, it is important to note that in countries where both freshwater and seawater are scarce, the approach employed in our study may overestimate the potential of hydrogen. Last but not least, compared to the energy systems literature [31-33], we incorporate country-specific discount rates in our calculations to account for the varying financial costs across countries.

Our results show that some countries possess both the physical and financial potential to meet their domestic energy needs and have the potential to further export significant amounts of renewable energy at a low cost (Figure 3). For instance, the analysis suggests that the US and China, currently the top two oil-importing countries [50], have an export potential of renewable energy at a production cost below 35 \$/MWh around 20 PWh, respectively. For context, this number is comparable to the current electricity consumption for the entire world [37]. Other countries that we identify as potential exporters include Saudi Arabia, Brazil and Kazakhstan. Our results regarding potential large exporters are in line with the main findings of Hjeij et al. [28] and the IRENA report [34], where big countries such as the US, China, Australia and Saudi Arabia emerge as the top contenders in the future hydrogen market. Australia does not rank within the top five countries in our study because we exclude remote sites located more than 200 km away from regions with grid access. This could change if Australia were to extend its grid to enable the development of these resources for export. Our results also indicate that some countries, such as Japan, South Korea, Belgium, the Netherlands and Germany, may not be self-sufficient in renewable energy or can do so only at an extremely high cost. Such conditions could prompt them to explore other low-carbon technologies, like nuclear power and CCS, or encourage them to import from low-cost countries with substantial potential. Even for countries that are potentially self-sufficient, but at a relatively high cost[3], importing electricity or electricity-derived fuels may be an attractive option, thereby stimulating trade. The ability of most countries to meet domestic demands with domestic resources offers them a choice between low-cost imports or securing higher-cost domestic supplies, which represents a marked contrast from the fossil economy, where energy self-sufficiency is impractical for most countries [51]. In a renewable future, some countries that are now large importers of fossil fuels can become exporters. Such shifts in energy imports and exports may have implications for foreign policy in the US [52] and China [53]. It is an open question whether the significant potential held by these two giants in comparison to the rest of the world (Figure 3) would result in a dominant position in the energy trade, akin to the current dominance of countries rich in oil resources.

In addition to the identification of potential importers and exporters, our findings also underscore the significant impacts of heterogeneous financial costs on both $RLCOE_{Ex}$ and $PEEV$. These results highlight the need to address financial obstacles, rather than technical challenges, to facilitate the development of renewable energy in certain countries effectively. For example, using North Africa to tap solar resources, as previously suggested by Refs. [11, 12], is perhaps not as economically attractive as the solar radiation data alone might suggest. Our findings regarding the impacts of heterogenous financial costs are consistent with the conclusions of Refs. [54-57], where renewable energy is less competitive in high-risk countries (Figures 6&S6).

In this study, we remain purposefully agnostic about future energy trade relationships. While our primary focus is on the potential for energy cost arbitrage, it's important to note that trade relationships are influenced by many other factors, including the availability of trading infrastructure (such as pipelines or import/export terminals), the cost of the trade itself and geopolitical relationships. The differences in renewable potential, here embodied using the $RLCOE_{Ex}$ and $PEEV$ metrics, between different countries create incentives and opportunities for international energy trade, but we acknowledge that this information alone does not predict what future trade patterns might look like. The energy trade, either through transmission grids or via some electricity-derived fuels such as ammonia, will come with a cost for transport, which is not included in our analysis. Therefore, the gap between one country's self-sufficiency margin and the export cost from another country ($RLCOE_{Ex}$), will partly be narrowed by the transport cost. Although we have not explicitly evaluated this cost, it seems that, at least for electricity, heterogeneity in resources does result in trade being an efficient way to decrease electricity system costs, at least within continents (by 10-30% [6, 58-60])[4]. As for hydrogen, a recent study conducted by Hampp et al. [61] showed that, even when factoring in transport costs, it is still more cost-effective to import hydrogen, methane, methanol and ammonia produced from other countries than domestically producing them in Germany. We also conducted a simple calculation of the levelized cost of hydrogen exported to Germany from Spain, Morocco and Saudi Arabia based on the cost parameters outlined in Hampp et al. [61]. The levelized cost of hydrogen exported from Spain, Morocco and Saudi Arabia is 21%, 19 % and 16% cheaper, respectively, compared to domestic hydrogen production in Germany. This example serves as further evidence that the heterogeneity of energy costs between countries can create an incentive for international energy trade. Notably, existing hydrogen trade agreements between Chile and Australia with Germany, Japan and the Netherlands [62] further highlight the growing trend of countries engaging in energy trade to capitalize on energy cost disparities.

[3] Around 30% of the countries display a $RLCOE_{Ex}$ above 30 $/MWh, while the global demand may be satisfied at a $RLCOE_{Ex}$ of 18 $/MWh (see Figure S5).

[4] As the distances become even longer, this effect diminishes. For instance, the reduction in electricity system costs resulting from intercontinental electricity trade is less than <5% [32,33].

We do not explore the impact of land availability for wind and solar on $RLCOE_{Ex}$ and $PEEV$. Lower land availability has an impact that is comparable to a higher electricity demand, as discussed in Results 2.4. For the country-specific discount rate, we use a value that reflects the present financial risk level for each country. The discount rate is particularly high for countries that experience political and social unrest today, which explains the high values in, e.g., Venezuela and Sudan (see Figure S7). However, these values can change. Countries that are unstable today may become stable in the future, while countries with a low-risk premium today may fall into socio-political unrest. We acknowledge this uncertainty and, thus, agree with the critique in Bogdanov et al. [63] about whether today's risk premiums accurately reflect a 2050 world. Addressing this uncertainty may require multiple scenarios of political stability and financial risk. Although outside the scope of this work, we welcome such efforts for future research. Furthermore, the discount rate varies not only between countries but also across different sectors, technologies and projects [64, 65]. Due to the project-specific nature of renewable energy technologies, we acknowledge the extreme difficulty of estimating the discount rates for every technology in each country. Instead, our focus is directed towards assessing how regional disparities in discount rates affect the potential for national renewable energy exports.

We argue that the $RLCOE_{Ex}$ and $PEEV$ metrics provide a solid basis for further scientific analysis of trade relations, thus removing the necessity for researchers to employ complex energy system models for such analysis. In addition, this work highlights the importance of a country's financial circumstances, alongside the renewable resource endowments, for its self-sufficiency and export potential of renewable energy. While we leave it to others to go deeper into the possible geopolitical scenarios that may be the consequence of the renewable energy cost reality, we hope that this paper can provide the basis for analyzing self-sufficiency and trade patterns for renewable electricity and electricity-derived fuels.

# 4. Methods

The approach for assessing the comparative competitiveness of countries in renewable energy exports consists of two parts: introducing the $RLCOE_{Ex}$ and $PEEV$ metrics and validating them using an energy system model. The development of the $RLCOE_{Ex}$ and $PEEV$ metrics involves two fundamental aspects: sourcing and processing data concerning wind, solar and hydropower output (Section 4.1), and assessing the financial costs linked to renewable energy in each country (Section 4.2). For details about validating the $RLCOE_{Ex}$ and $PEEV$ metrics, an introduction to the energy system model and the cost assumptions for various energy technologies, please refer to Section 4.4.

## 4.1 Renewable *LCOE* (*RLCOE*)

We first calculate the cost of supplying one unit of wind or solar energy from a grid cell 0.01° × 0.01° (approximately 1 km × 1 km at the equator) based on the ERA5 reanalysis data (hourly wind speed, direct and diffuse solar insolation) [66], annual average wind speed from Global Wind Atlas (GWA) [67] and a uniform discount rate of 5%. The $RLCOE$ for global onshore wind and solar resources is illustrated in Figure S6.

**Wind capacity factor**

The wind profile, which represents the hour-to-hour variation in wind speed, and the annual average wind speed are the two key parameters for assessing the wind power potential. The ERA5 data provides an accurate estimate of the wind profile [68]. However, its low spatial resolution (31 km × 31 km) means that it is not well-suited for assessing the annual average wind speed given the potential heterogeneity in wind speed within a small geographical area [69]. The annual average wind speed is accurately documented in the GWA [67]. Thus, we combine the ERA5 data set (wind profile) and the GWA dataset (annual average wind speed) with the methodology from Mattson et al. [69]. Each small pixel (with a size same as that in GWA, 1 km × 1 km) is provided with the wind profile from the corresponding larger pixel in ERA5, and the wind profile is then scaled using the average wind speed in GWA. By doing so, we obtain an hourly time series of wind speed that captures geographical variations in wind output caused by local differences in topography and land cover at a spatial resolution of 1 km (compared to 31 km for ERA5). The instantaneous wind speeds are then converted into capacity factors using the output profile of the 3 MW Vestas V112 wind turbine, including wake losses and Gaussian smoothing to account for wind variations within a park (Figure S8) [69]. The annual mean wind power capacity factor is calculated by averaging the hourly wind power capacity factor over one year.

**Solar capacity factor**

The solar capacity factor is estimated based on the ERA5 "surface solar radiation downwards" (SSRD) and "total sky direct solar radiation at surface" (FDIR) [66]. In addition to these two ERA5 variables for diffuse and direct insolation, we also need top-of-atmosphere solar insolation (TOA) variations over the year. This variable is calculated as below [70]:

$$\mathrm{TOA} = I_0(1 + 0.034cos\frac{2\pi n}{365.25})$$

where $I_0$ is the solar constant (1361 W/m2) and $n$ is the ordinal of the day in the year.

The total insolation striking a tilted solar PV panel is the Global Tilted Irradiance (GTI):

$$\mathrm{GTI} = I_{direct}^{sun} + I_{diffuse}^{sky} + I_{diffuse}^{ground}$$

where $I_{direct}^{sun}$ is direct beam radiation from the sun, $I_{diffuse}^{sky}$ is diffuse radiation from the sky and $I_{diffuse}^{ground}$ is diffuse reflected radiation from the ground. $I_{direct}^{sun}$ can be directly calculated from the ERA5 FDIR variable using the solar position. $I_{diffuse}^{ground}$ is also straightforward assuming a constant uniform ground albedo. We use the Hay-Davies model which includes an isotropic component and circumsolar diffuse radiation to take into account that the sky is brighter nearer to the sun. The resulting equations are:

$$I_{direct}^{sun} = \mathrm{FDIR} \cdot R_b = \mathrm{FDIR} \cdot \frac{\cos \mathrm{AOI}}{\cos z} = \mathrm{DNI}.\cos AOI$$

$$I_{diffue}^{sky} = \mathrm{DHI} \cdot \mathrm{AI} \cdot R_b + \mathrm{DHI} \cdot (1 - \mathrm{AI}) \cdot \frac{1 + \cos \beta}{2}$$

$$I_{diffue}^{ground} = \mathrm{GHI} \cdot \rho \cdot \frac{1 - \cos \beta}{2}$$

where $R_b$ is the ratio of tilted and horizontal solar beam irradiance, AOI is the angle of incidence of the sun on the PV panel, $z$ is the solar zenith angle, DNI is direct normal irradiance, DHI is diffuse horizontal irradiance, AI is the anisotropic index (a measure of nonuniformity of sky brightness), $\beta$ is the tilt angle of the PV panel and $\rho$ is ground albedo, which is assumed to be 0.2 everywhere. The variables are further related by:

$$\mathrm{DHI} = \mathrm{SSRD}, \qquad \mathrm{DNI} = \frac{\mathrm{FDRI}}{\cos z}, \qquad \mathrm{AI} = \frac{\mathrm{DNI}}{TOA}, \qquad R_b = \frac{\cos \mathrm{AOI}}{\cos z}$$

$$\cos \mathrm{AOI} = \cos z \cos \beta + \sin z \sin \beta \cos(\alpha_{sun} - \alpha_{PV})$$

Here $\alpha_{sun}$ is the azimuth angle of the sun and $\alpha_{PV}$ is the azimuth angle of the PV panel (assumed zero), with azimuth measured with zero due south and positive direction toward west. ERA5 radiation variables are documented in Hogan [71].

In clear-sky weather, the optimal tilt angle of a PV module for a given location is the latitude of the panel. However, if conditions are often cloudy, more diffuse sky radiation can be captured if the tilt angle is smaller than its latitude. Therefore, the optimal tilt angle is location specific. For simplicity, we use the fitted third degree polynomials from Jacobson et al. [72] to get near optimal tilt as a function of latitude and we do not consider tracking solar PV systems.

Given that solar radiation is rather stable within a certain geographical area (compared with the heterogeneity in wind speed), the calculated solar capacity factor based on ERA5 for each large pixel (31 km) is then provided to the corresponding small pixels (1 km). In this way, we get a map for the solar capacity factor with the same resolution as the wind capacity factor.

**Cost assumptions**

The costs for wind and solar in 2050 are based on the estimates from IRENA [73, 74]. We do not explicitly consider learning rates, meaning that we do not have endogenous learning in the analysis. Instead, we assume that the cost declines implied in the IRENA estimates for 2050 adequately capture the combined effects of local and global learning for a level of deployment sufficient to achieve a 100% renewables-based energy system. For more discussion about cost assumptions, please refer to Supplementary Information 2.1. All the cost assumptions and technical parameter values are summarized in Table 1. Note that most utility-scale PV and onshore wind projects do not own the land on which the PV panels and wind turbines are placed. The land lease cost consists of a minor share of the total cost for the solar and wind power project [75, 76].

**Table 1** Cost data and technical parameters.

| Technology | Investment cost [$/kW] | Variable O&M costs [$/MWh] | Fixed O&M costs [$/kW/yr][a] | Lifetime [years] |
|---|---|---|---|---|
| Onshore wind | 825 [b] | 0 | 33 | 25 |
| Offshore wind | 1500 [b] | 0 | 55 | 25 |
| Solar PV | 323 [c] | 0 | 8 | 25 |
| Solar rooftop | 423 [c] | 0 | 6 | 25 |

[a] Akar et al. [77]

[b] IRENA [73]

[c] IRENA [74]

## 4.2 Renewable *LCOE* with country-specific discount rate

The $RLCOE$ is first calculated with a uniform discount rate of 5% for the entire world, similar to the common practice in other studies, see Figures S6a&c. By contrast, **Renewable *LCOE* with country-specific discount rate** ($RLCOE_r$) takes into account the different circumstances for investment in different countries, see Figures S6b&d. The fixed investment costs for renewable power plants can be characterized using an overnight capital cost (OCC), potentially modified by a cost of capital during construction, which is depreciated over the economic lifetime of a project using a weighted average cost of capital (WACC). Both OCC and WACC can vary regionally. For instance, the IEA's World Energy Outlook 2021 used 600 $/kW as OCC for solar PV in India and 1100 $/kW in the US [78]. Additionally, it applied a WACC of 3% to 6% for solar PV and onshore wind projects [78]. OCC can even vary significantly within a single country. In the Annual Energy Outlook 2023, the US EIA employed OCC for onshore wind that varies from 1566 to 3458 $/kW across the 25 regions modeled in the United States [79].

OCC includes the costs of materials, equipment and labor, and can also include the cost of land acquisition, grid interconnection, permitting and other professional services. Regional differences can be driven by the costs of both skilled and unskilled labor, remoteness of the site and the regulatory environment in which a project is developed, among other factors, each of which can vary over time as well. WACC incorporates the financing structure of a specific project, including the costs of equity and debt financing, along with any government support, such as guarantees, subsidies, favorable tax, royalty treatment or direct financial contributions. These items can vary from project to project, across companies and industries, and are dependent on the priorities of national and local governments, which can sometimes change abruptly.

In the present study, we emphasize the impacts of regional differences in levelized capital costs on the deployment potential of wind and solar energy, as well as the corresponding impacts on the national competitiveness of renewable energy exports. Levelized capital costs can vary due to differences in resource quality, capital costs or WACC. Due to the project-specific nature of many of the capital cost drivers, we recognize the futility of trying to estimate average capital costs from a bottom-up analysis of their constituent components. Instead of estimating regional OCC and WACC separately, we take a different, top-down approach to estimate differences in levelized capital costs. Our approach modifies a uniform capital cost baseline using a country-specific hurdle rate that captures the overall difficulty of doing business in a country. We are unaware of any comprehensive global studies about the country-specific variations in wind and solar PV capital costs. In addition, we anticipate that idiosyncratic capital cost differences across individual renewable energy projects would become less pronounced under the type of large-scale building program encompassing many individual projects that would be required under a low-carbon energy transition. We therefore assume for the sake of illustration that the levelized capital cost differences for wind and solar are dominated primarily by the quality of the resource and the cost of capital (i.e., WACC). We assume a common capital cost (unmodified OCC) for all projects, and levelize it over the lifetime of the asset using country-specific discount rates that incorporate risk premium estimates from Damodaran [38]. These estimates are available for most countries and are given in the form of an additional hurdle rate above a common global risk-free yield. They are based on objective financial measures (e.g., credit default swap spreads from sovereign bond yields), where available, and subjective sovereign credit risk ratings from Moody's or Standard & Poor's where government bonds are not widely traded [38]. While such country risk premiums technically correspond only to sovereign default risk, the ability of a government to support a multi-decadal, large-scale infrastructure program depends on many of the same drivers, such as macroeconomic and political stability. We add the country risk premiums to the uniform discount rate (5%) to obtain the country-specific discount rates that are employed in calculating the $RLCOE_r$ for each country in this study. Note that the financial costs of an individual project may differ significantly from the country-specific discount rates. For more discussion about the individual project's financial costs, please refer to Supplementary Information 2.2.

### 4.3 Renewable Levelized Cost of Energy available for Export ($RLCOE_{Ex}$) and Potential Energy Export Volume (*PEEV*)

We estimate the **Renewable Levelized Cost of Energy available for Export** ($RLCOE_{Ex}$) for most countries in the world. It measures the marginal cost for a country to supply its entire energy demand using only domestic renewable resources (wind, solar and existing hydropower) (Figure 1). Here, the energy demand includes both electricity demand and hydrogen demand. Our initial step involves arranging the $RLCOE_r$ in ascending order to construct the national renewable energy supply curve (Figure 1). Subsequently, we pinpoint the annual domestic energy demand marked by the red line at 1.0 on the x-axis in Figure 1. The point of intersection between the annual supply and annual demand on the supply curve signifies the $RLCOE_{Ex}$. For instance, if a country has an annual energy demand of 100 TWh and an annual hydropower generation of 20 TWh, the $RLCOE_{Ex}$ is determined by sorting $RLCOE_r$ for all the grid cells within that country until the total generation reaches 80 TWh. $RLCOE_{Ex}$ corresponds to the $RLCOE_r$ of the last grid cell required to achieve a generation equal to the annual demand.

$RLCOE_{Ex}$ is estimated at the country level or subnational level for some big countries. Figure S9 illustrates the national renewable energy supply curves for a random selection of countries. Remote solar and wind power plants that are far from regions with grid access may require additional investments in transmission grids. Therefore, we add 200 $/kW as extra investments in transmission grids for remote solar PV and wind power plants [69]. As for the hydropower potential, we obtain the data from Refs. [80-82]. The $RLCOE_{Ex}$ thus assesses the renewable energy potential in relation to the energy demand of the country. It is a metric that hints at the national self-sufficiency potential (if $RLCOE_{Ex}$ < certain reasonable cost), as well as the export potential (If $RLCOE_{Ex}$ is very low, there is likely an export potential of electricity or electricity-derived fuels). It takes into account, in addition to the country-specific discount rate, the available land for renewable energy in relation to the domestic energy demand.

Taking into consideration a threshold on the national renewable energy supply curve, it is possible to estimate the overall renewable energy production (*Pt*) within the specified threshold (Figure 1). By deducting the domestic energy demand from *Pt*, we derive the **Potential Energy Export Volume** (*PEEV*, in PWh) (Figures 1&3). The *PEEV* metric serves as an approximation of the amount of renewable energy that could be economically produced and traded in different countries in a renewable future. For more discussions about the methodology, please refer to Supplementary Information 2.3.

It is important to note that all renewable energy resources, including wind, solar, hydro, biomass and geothermal, along with other power generation technologies like nuclear power and CCS, can be integrated as energy supply technologies for the estimation of $RLCOE_{Ex}$ and $PEEV$. The reason why we concentrate on wind and solar for energy supply is twofold. First, these resources are abundant, widely available, and the technologies harnessing them are becoming increasingly cost-competitive compared to other power generation methods. Second, our primary goal is to elucidate concepts rather than predict eventual winners among technologies. For simplicity, we primarily focus on wind and solar energy, along with existing hydropower. Additionally, both the $RLCOE_{Ex}$ and $PEEV$ metrics are agnostic to the nature of future energy trade. These metrics serve as generic indicators for evaluating energy trade, irrespective of the energy carrier. For a more detailed discussion on the format of future energy trade, please refer to Supplementary Information 2.4.

**Energy demand**

The annual electricity consumption for each country in 2050 is estimated by extrapolating the annual demand in 2016 [83]. This extrapolation is based on the regional demand growth between 2016 and 2050 in the Shared Socioeconomic Pathway 2 scenario outlined in the IPCC report [14]. We then estimate the hourly demand profile based on a machine learning approach which adopts historical demand profiles for 44 countries as input to a gradient boosting regression model [84] to calculate the hourly demand profile. The regression model takes into account the calendar effects (e.g., hour of day, weekday and weekend), temperature (e.g., hourly temperature in the most populated areas of each region), and economic indicators (e.g., local GDP per capita). Finally, the hourly demand series is scaled to match the annual electricity demand for each region in 2050. As for hydrogen demand, we assume the annual demand for hydrogen equals half the annual electricity demand, which is consistent with the magnitude of projected hydrogen demand for 2050 outlined in the European Commission's long-term strategic vision [85].

The annual electricity demand and the annual hydrogen demand are combined to form the domestic energy demand used in the calculation of $RLCOE_{Ex}$ and $PEEV$. Meanwhile, the hourly electricity demand profile and the annual hydrogen demand are utilized as inputs for an energy system model, which is employed to calculate the marginal hydrogen cost and the potential hydrogen export volume (see Methods 4.4). For sensitivity analysis, we double the electricity demand to account for large-scale electrification, which is consistent with the estimations from sector-coupling energy system studies [3, 86]. For simplicity, we assume that the energy demand is inelastic. Please refer to Supplementary Information 2.1 for more discussion about energy demand.

**Assumptions on land availability for wind and solar**

A crucial parameter needed to estimate the $RLCOE_{Ex}$ is how densely solar and wind power may be deployed in the landscape, and which types of land to exclude from potential wind and solar exploitation. Many different assumptions are made in the literature for wind power [23], and there is sparse empirical evidence for those assumptions [87]. The analysis in Hedenus et al. [87] suggests that wind turbines have been built on all kinds of land types, and up to 20% of all the land has been used for wind deployment in some counties in the US. Since institutional frameworks differ between countries, ideally, assumptions regarding restrictions on where to deploy solar and wind power should be dependent on each country. However, as such analyses have not yet been done, we here simply assume that wind power may be deployed on all types of land, but that a maximum of 10% of the land may be exploited for wind power purposes. As for offshore wind power, we assume it can be installed in areas with a maximum depth of less than 60 m, and a maximum of 10% of the area may be deployed for wind power. Given the limited knowledge about where and how much solar PV may be built, we make more conservative assumptions for solar PV. We exclude all land covered with forests and assume a maximum of 5% of the remaining land to be available for solar PV installations. For Rooftop solar PV, we assume that 5% of the urban areas can be utilized for its installations.

**Table 2** Assumptions made regarding the capacity limits of wind and solar PV.

| | **Solar PV** | **Solar Rooftop** | **Onshore wind** | **Offshore wind** |
|---|---|---|---|---|
| Density [$W/m^2$] [a] | 45 | 45 | 5 | 5 |
| Available land [%] | 5% | 5% | 10% | 10% |

[a]The term 'Density' refers to the capacity assumed to be installed per unit area for a typical solar or wind farm.

## 4.4 Marginal hydrogen cost and potential hydrogen export volume

The $RLCOE_{Ex}$ and $PEEV$ metrics focus only on the costs of power generation. The cost of energy in a renewable energy system consists of both generation costs and the costs to manage the variation of wind and solar [6]. To validate the comparative competitiveness of exports based on $RLCOE_{Ex}$ and $PEEV$ analysis and findings regarding potential importers and exporters, we employ a typical techno-economic cost optimization model (Supergrid) to investigate each country covered in our study [41]. The Supergrid model is a greenfield capacity expansion model with hourly time resolution, which optimizes the investment and dispatch for the electricity sector and hydrogen production with an overnight approach. The exception is hydropower, where existing hydropower plants are assumed to be still in operation in 2050 and the capacity is assumed to remain at the current level due to environmental regulations. In terms of the $CO_2$ emission target, we assume a nearly zero emission system with a global $CO_2$ emission cap of 1g $CO_2$ per kWh of energy demand. The model is written in the Julia programming language using the JuMP optimization package. The model-specific code, input data and output data are available online to further enhance the transparency and reproducibility of the results. The cost assumptions and key parameters for technologies are summarized in Table 3. For a more detailed description of the model, see Ref. [41].

We first calculate the marginal hydrogen cost for each country. The marginal hydrogen cost refers to the shadow price of the hydrogen balancing constraint. The marginal hydrogen cost represents the cost of producing an additional unit of hydrogen for export after the domestic energy demand has been met. Unlike the simple $RLCOE_{Ex}$ metric, the marginal hydrogen cost represents the cost for hydrogen at the export node. By comparing the $RLCOE_{Ex}$ results with the marginal hydrogen costs, we can assess whether our conclusions about the relative national competitiveness in exporting renewable energy, based on $RLCOE_{Ex}$ analysis, remain valid when accounting for system integration costs and hydrogen production.

By conducting a thorough analysis of the marginal cost of hydrogen at various hydrogen demand levels, we can generate a hydrogen supply curve for each country (Figures S10 & S11). By setting a threshold on the national hydrogen supply curve, we can calculate the potential total hydrogen production achievable at the designated threshold. Subtracting the domestic hydrogen demand from this total hydrogen production allows us to determine the potential hydrogen export volume. We then validate the partition of countries into those with a large export potential from those with a potential import demand by comparing the $PEEV$ with the potential hydrogen export volume. For the validation, we choose the threshold value for the marginal hydrogen cost based on the trendline in Figure 4, where an $RLCOE_{Ex}$ value of 35 $/MWh corresponds to a marginal hydrogen cost of 60 $/MWh.

**Table 3** Cost data and technical parameters.

| Technology | Investment cost [$/kW] | Variable O&M costs [$/MWh] | Fixed O&M costs [$/kW/yr] | Fuel costs [$/MWh fuel] | Lifetime [years] | Efficiency/ Round-trip efficiency |
|---|---|---|---|---|---|---|
| Natural gas OCGT | 500 | 1 | 10 | 22 | 30 | 0.35 |
| Natural gas CCGT | 800 | 1 | 16 | 22 | 30 | 0.6 |
| Coal | 1600 | 2 | 48 | 11 | 40 | 0.45 |
| Biogas OCGT | 500 | 1 | 10 | 37 | 30 | 0.35 |
| Biogas CCGT | 800 | 1 | 16 | 37 | 30 | 0.6 |
| Onshore wind [a] | 825 | 0 | 33 | n/a | 25 | n/a |
| Offshore wind [a] | 1500 | 0 | 55 | n/a | 25 | n/a |
| Solar PV [b] | 323 | 0 | 8 | n/a | 25 | n/a |
| Solar Rooftop [b] | 423 | 0 | 5.8 | n/a | 25 | n/a |
| Electrolyzer | 250 | 0 | 5 | n/a | 25 | 0.66 |
| Hydrogen storage | 11 $/kWh | 0 | 0 | n/a | 20 | n/a |
| Fuel cell | 800 | 0 | 40 | n/a | 10 | 0.5 |
| Hydro | 300[c] | 0 | 25 | n/a | 80 | 1 |
| Onshore Transmission[d] | 400 $/MW/km | 0 | 8 $/MW/km | n/a | 40 | 0.016 loss per 1000 km |
| Offshore Transmission[e] | 470 $/MW/km | 0 | 1.65 $/MW/km | n/a | 40 | 0.016 loss per 1000 km |
| Converter[d] | 150 | 0 | 3.6 | n/a | 40 | 0.986 |
| Battery[f] | 116 $/kWh | 0 | 0 | n/a | 15 | 0.9 |

[a]IRENA [73]

[b]IRENA [74]

[c]Steffen [88], this cost pertains to the expenses linked to the replacement of old mechanical and electrical machinery.

[d]Hagspiel et al. [89]

[e]Purvins et al. [90]

[f]Cole et al. [91]

OCGT, Open-cycle gas turbine; CCGT, combined-cycle gas turbine.

## Code and data availability

The code and data supporting the $RLCOE_{Ex}$ and $PEEV$ metrics can be accessed via this link: https://github.com/xiaomingk/Global-renewable-potential. The code and data for the Supergrid model can be found at this link: https://github.com/xiaomingk/Supergrid.

# Supplementary Information

## Renewable levelized cost of energy available for export: An indicator for exploring global renewable energy trade potential

Xiaoming Kan[a,*] Lina Reichenberg[a] Fredrik Hedenus[a] David Daniels[a]

[a]Department of Space, Earth and Environment, Chalmers University of Technology, 41296 Gothenburg, Sweden

* Correspondence: kanx@chalmers.se

# 1. Supplementary figures

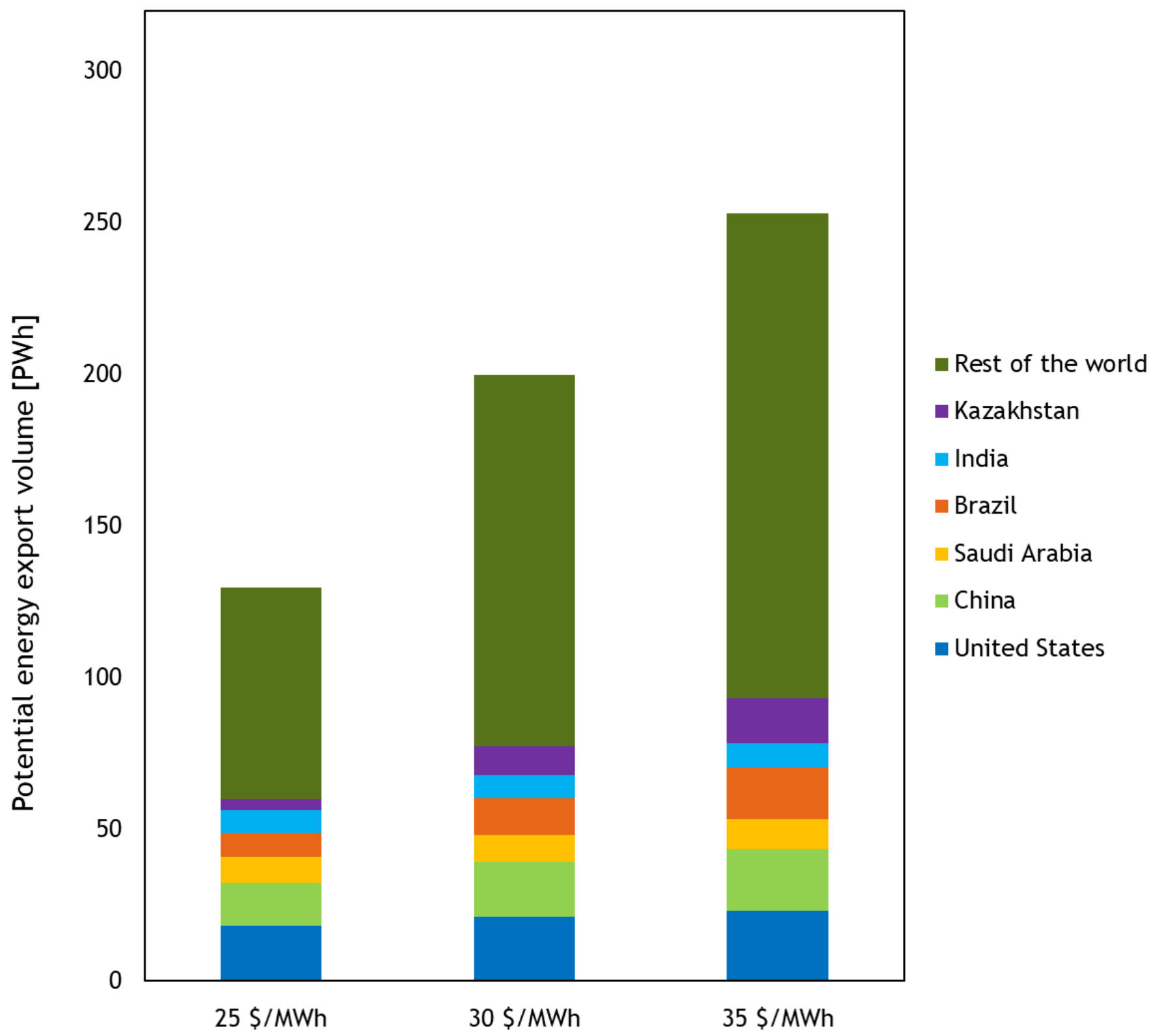

**Figure S1** Export potential under different assumptions regarding the threshold on the national renewable energy supply curve

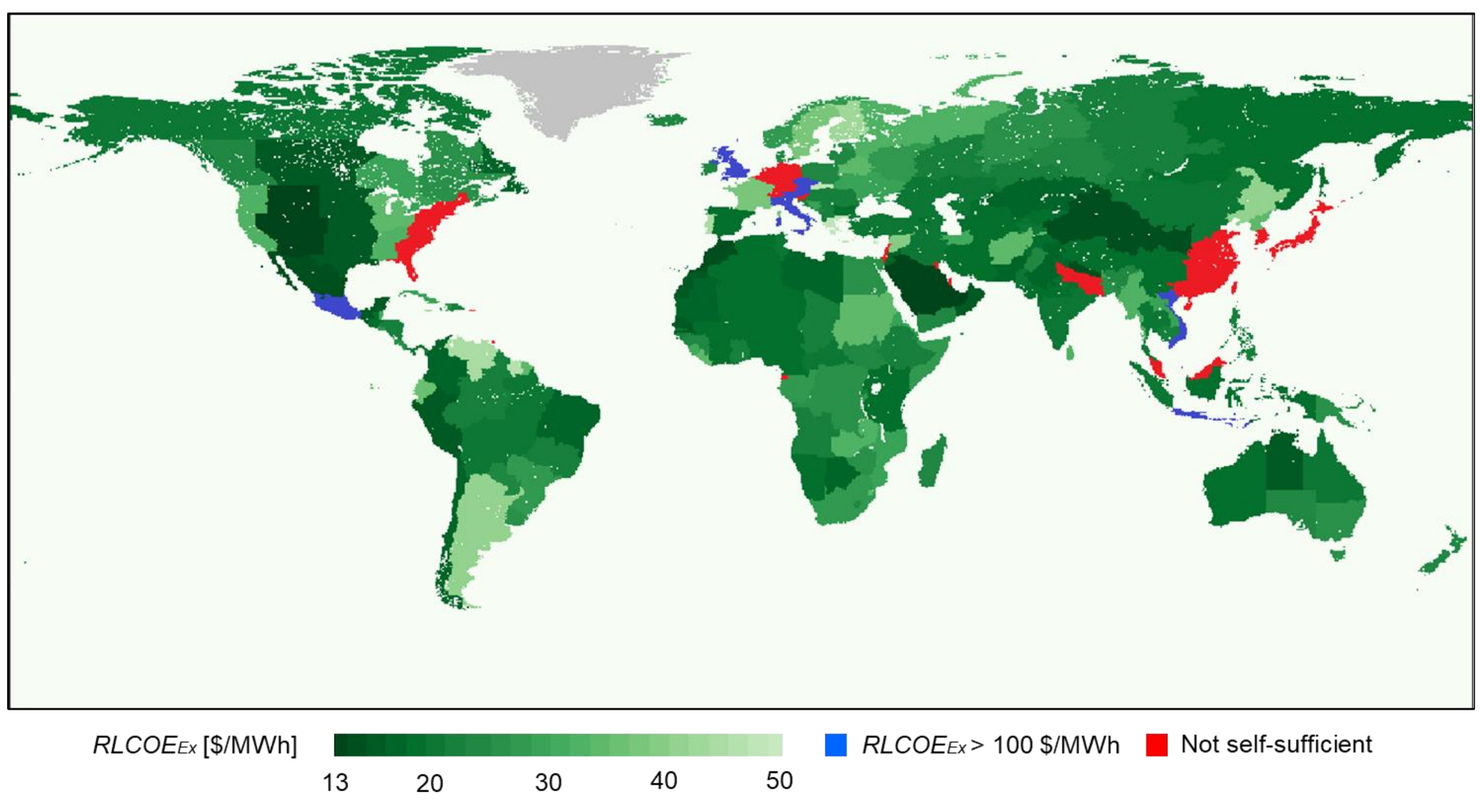

**Figure S2** *RLCOE*$_{Ex}$ for 2050 under the assumption of doubling electricity demand due to increased electrification

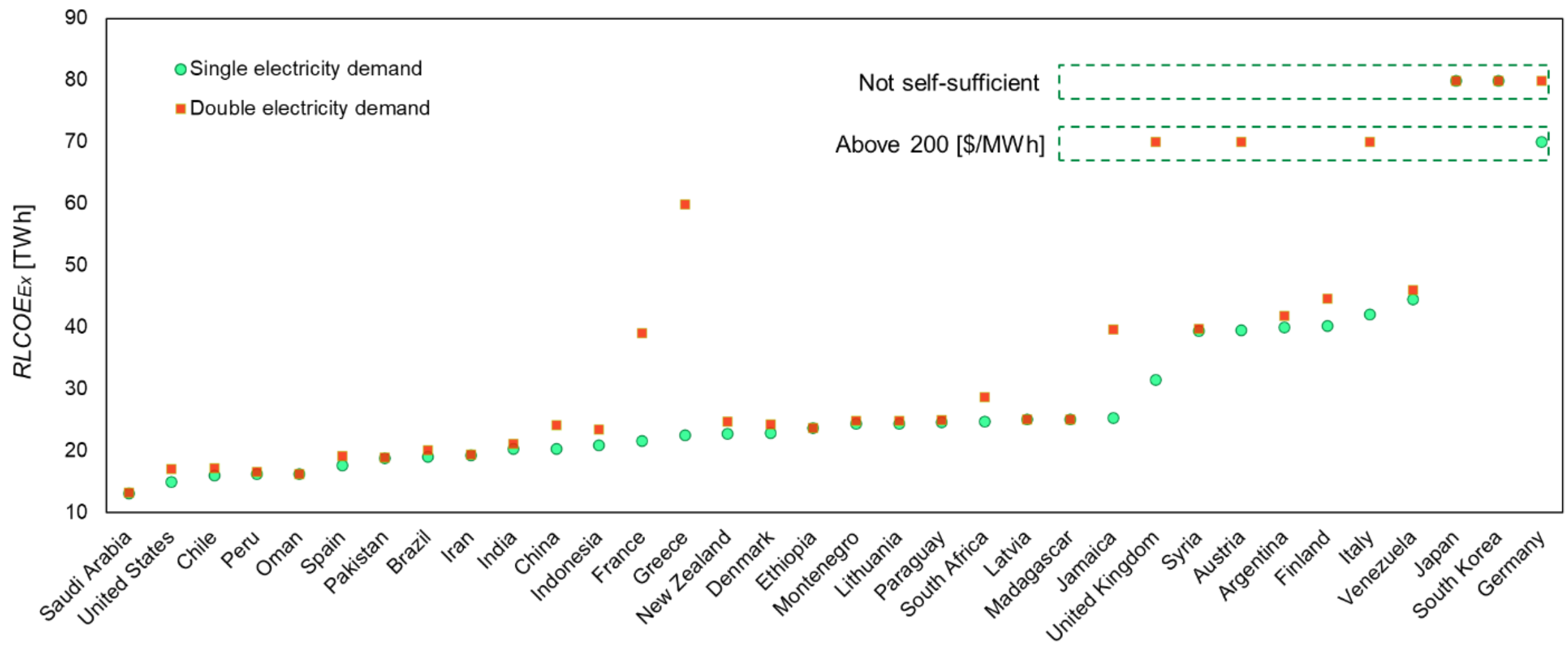

**Figure S3** $RLCOE_{Ex}$ for select countries at various electricity demand levels. Japan and South Korea are not self-sufficient regardless of the demand level, while Germany is not self-sufficient if electricity demand is doubled.

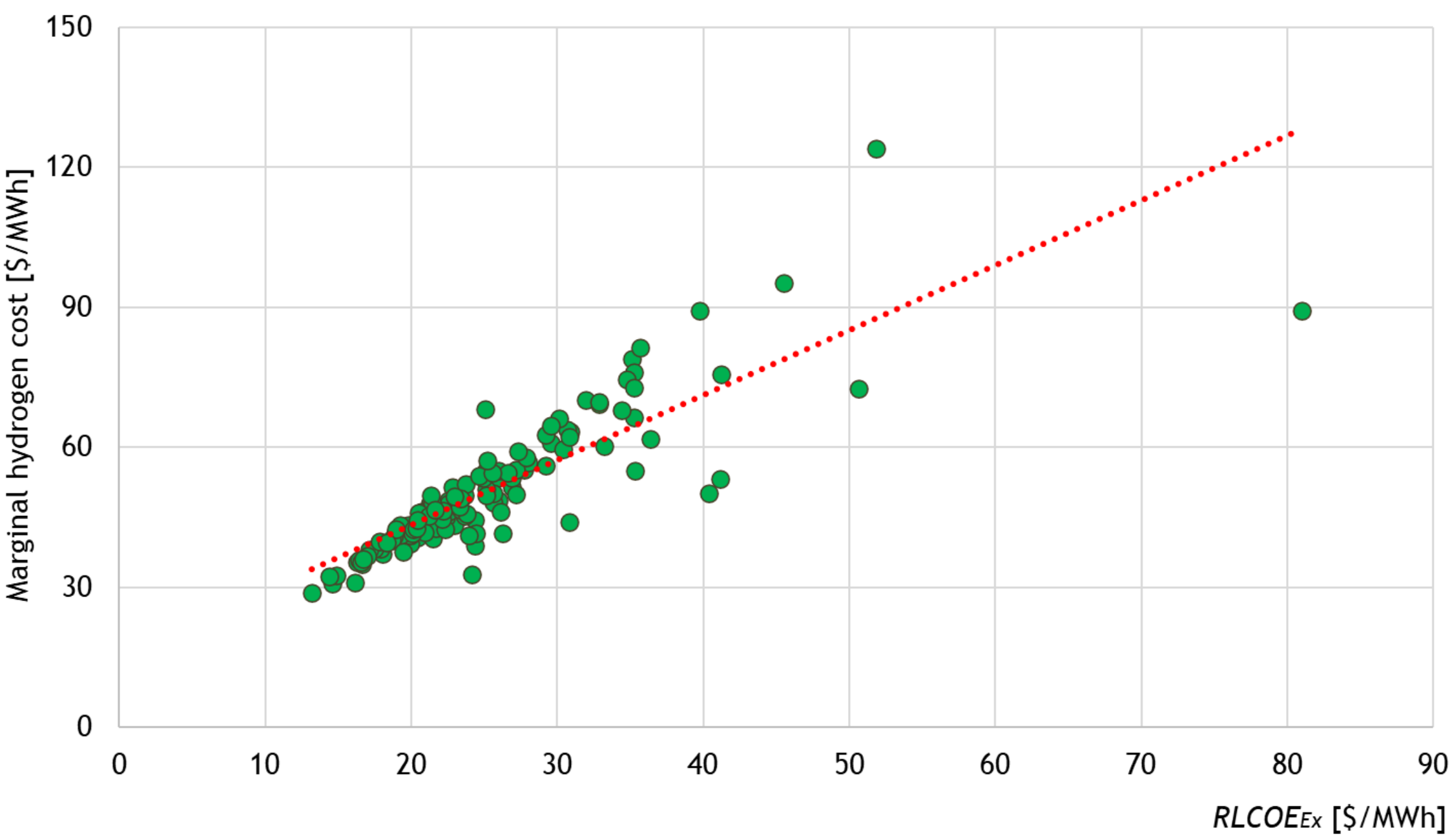

**Figure S4** Relationships between the marginal hydrogen cost and $RLCOE_{Ex}$ when the hydrogen demand is doubled. To enhance visualization clarity countries that are not self-sufficient or exhibit exceptionally high costs due to land constraints are excluded. Only countries/regions with a $RLCOE_{Ex}$ below 90 $/MWh are considered.

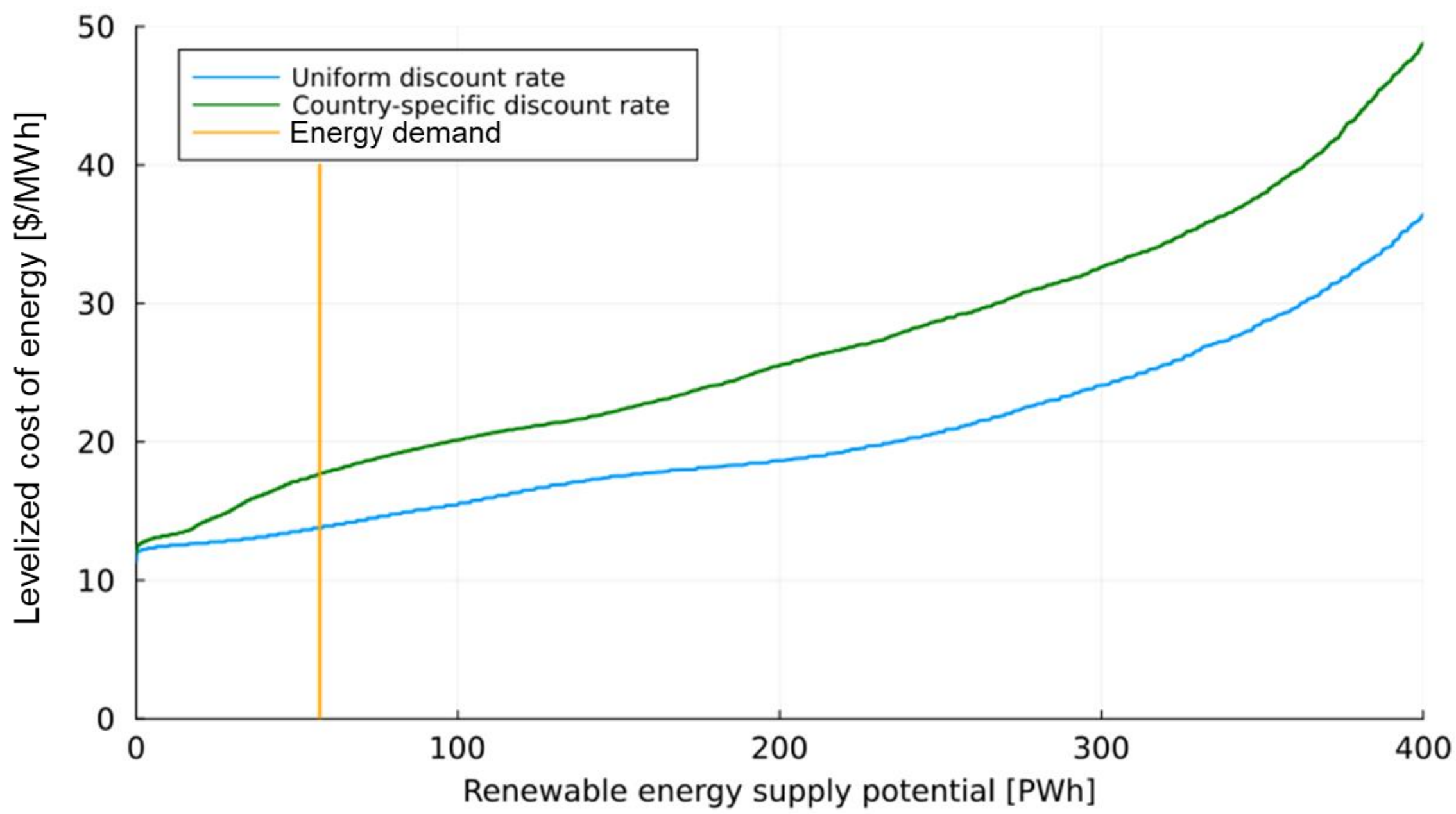

**Figure S5** Renewable energy supply curve for the entire world under different discount rates. The global renewable energy supply potential is abundant compared with the energy demand in 2050. Specifically, the global energy demand can be met at a cost of 18 $/MWh, which is 28 % more expensive than the value based on a uniform discount rate.

[$/MWh]
≥50
40
30
20
10

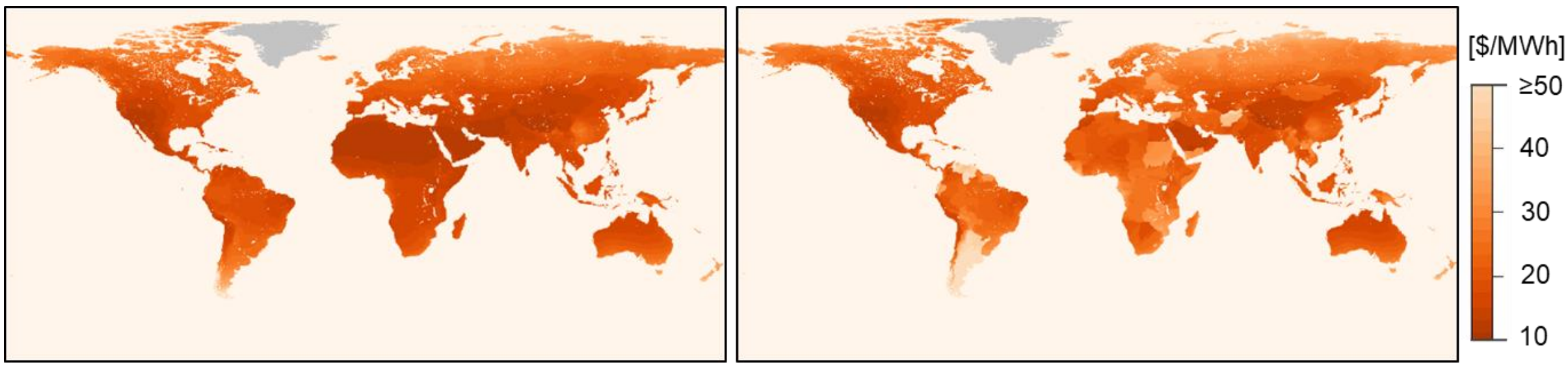

a) Solar *LCOE*_uniform discount rate

b) Solar *LCOE*_country-specific discount rate

[$/MWh]
≥50
40
30
20
10

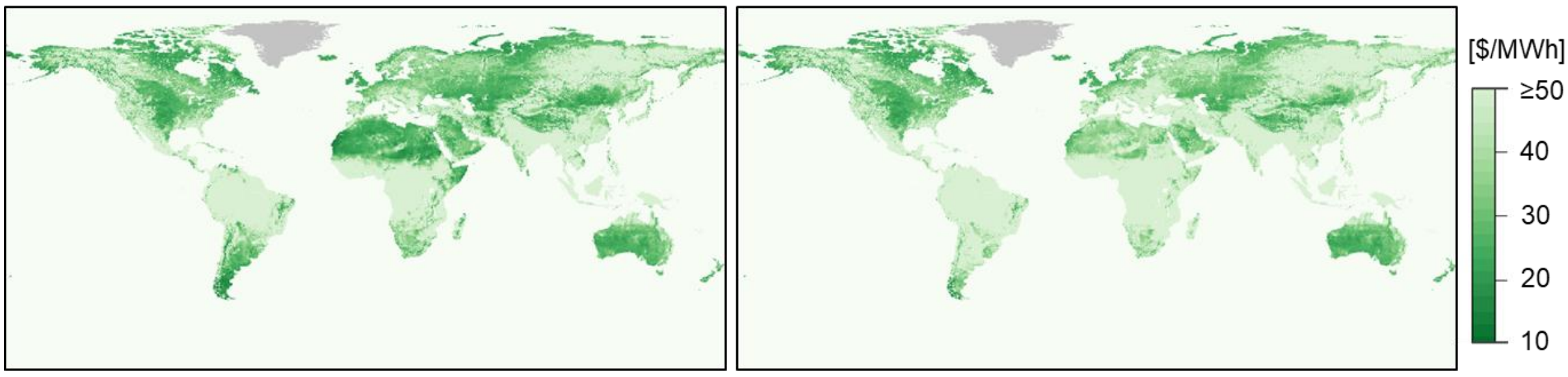

c) Wind *LCOE*_uniform discount rate

d) Wind *LCOE*_country-specific discount rate

[$/MWh]
≥50
40
30
20
10
5
1
0

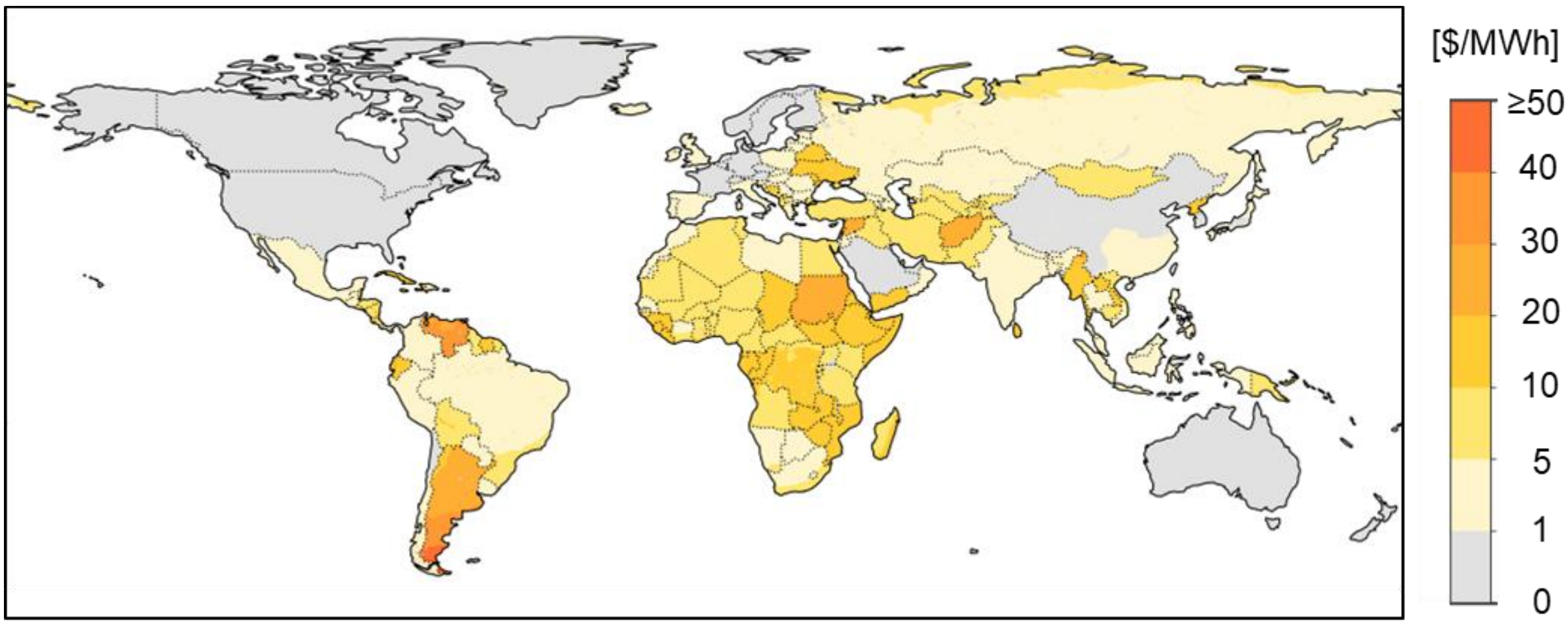

e) Solar *LCOE* difference between cases with a uniform discount rate and a country-specific discount rate

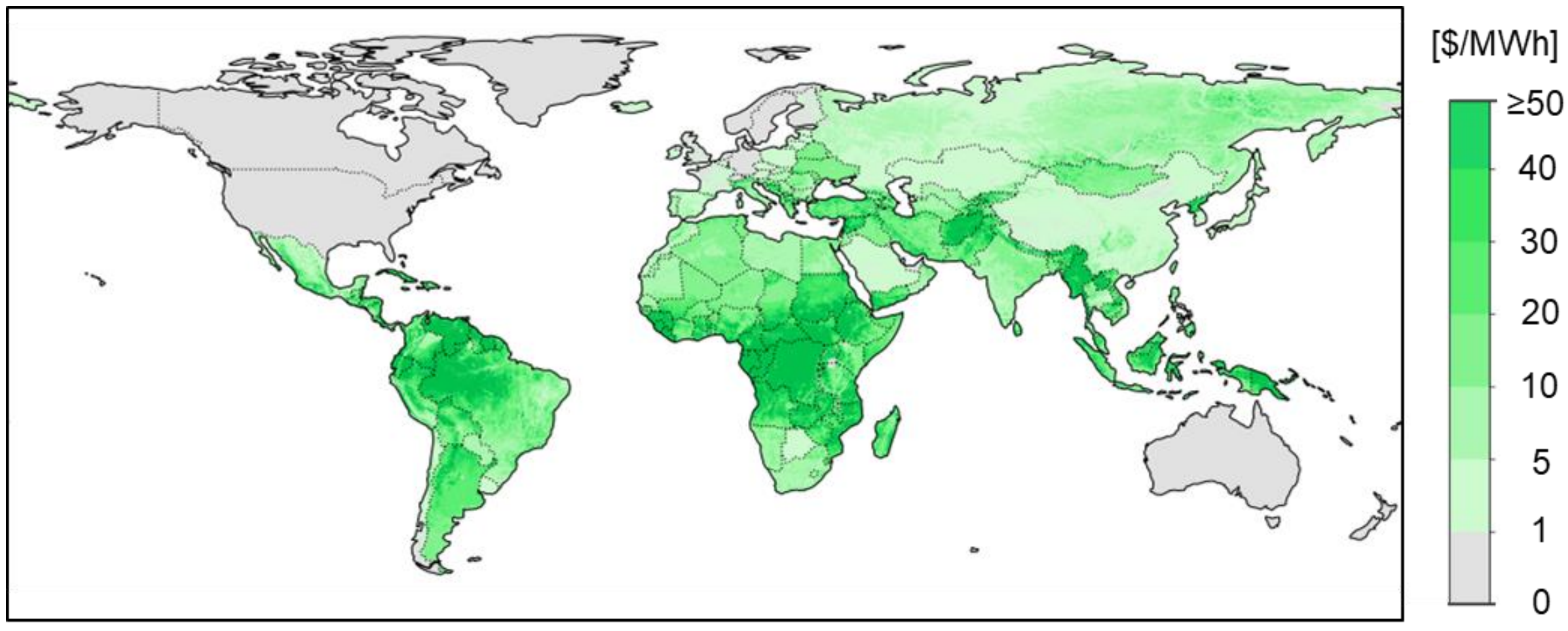

f) Wind *LCOE* difference between cases with a uniform discount rate and a country-specific discount rate

**Figure S6** The Renewable *LCOE* for solar PV and onshore wind power using cost assumptions for 2050 under uniform and heterogeneous discount rates. For Figures a and c, the *LCOE* is calculated for each grid cell (0.01° × 0.01°) using a uniform discount rate. The cheapest sites for solar PV can be found around the Tropic of Cancer

and in most parts of Africa. Some regions close to the equator have mediocre resources due to cloudiness, for instance, East Indonesia and Central South America. Applying the country-specific discount rates changes the *LCOE* for solar rather significantly, see Figure b). The lowest costs are now found in the US, West China, and Saudi Arabia. The cost for solar PV in Africa and Latin America turns out to be relatively high when the heterogeneity in financial costs is considered, comparable to Central or even Northern Europe, despite the latter having much less favorable solar conditions. The wind potential is more heterogonous compared to the solar potential. The cheapest supply is found in North Africa, Central Asia and Southern Latin America. By contrast, areas around the equator display relatively poor wind conditions. The *LCOE* for wind when country-specific discount rates are considered is shown in Figure d). The main difference compared to the case of uniform discount rate is that North Africa and Argentina are less attractive places to invest in wind power, whereas Central US, Australia and Central Asia still display a wind *LCOE* of around 25 $/MWh. Figures e and f illustrate the difference in solar *LCOE* and wind *LCOE* between the case with a uniform discount rate and the case with a country-specific discount rate, respectively.

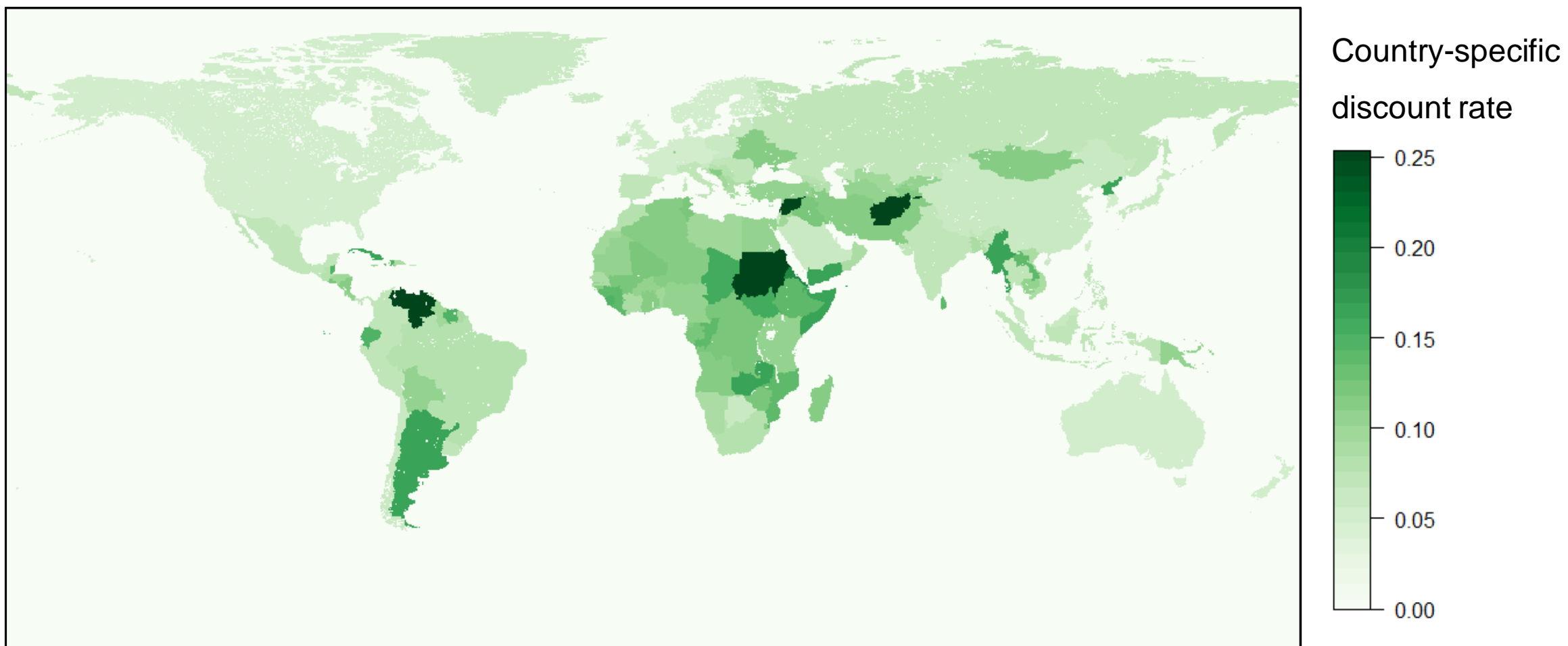

**Figure S7** Country-specific discount rates for most countries in the world

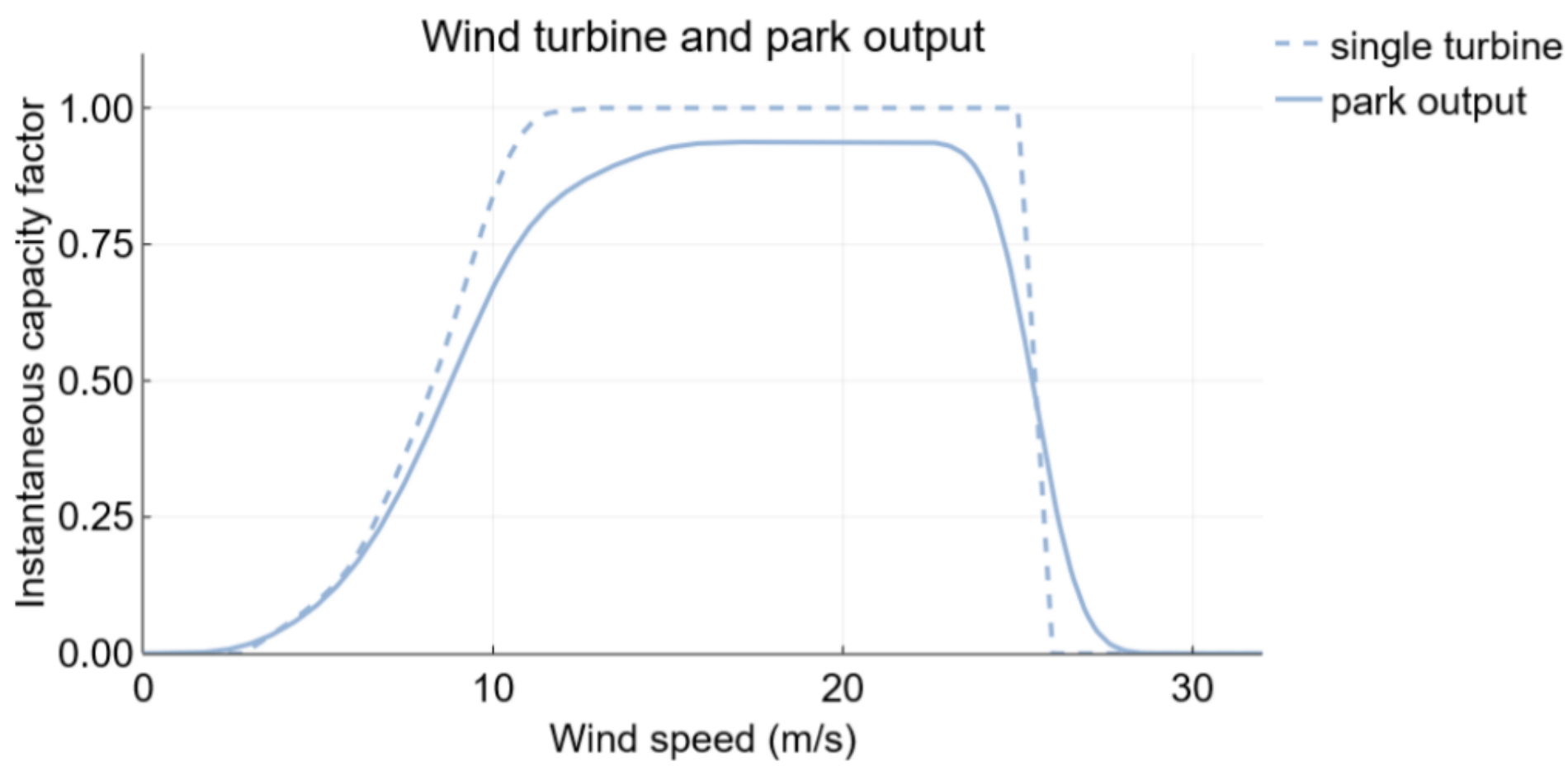

**Figure S8** The output profile of the 3 MW Vestas V112 wind turbine and wind park

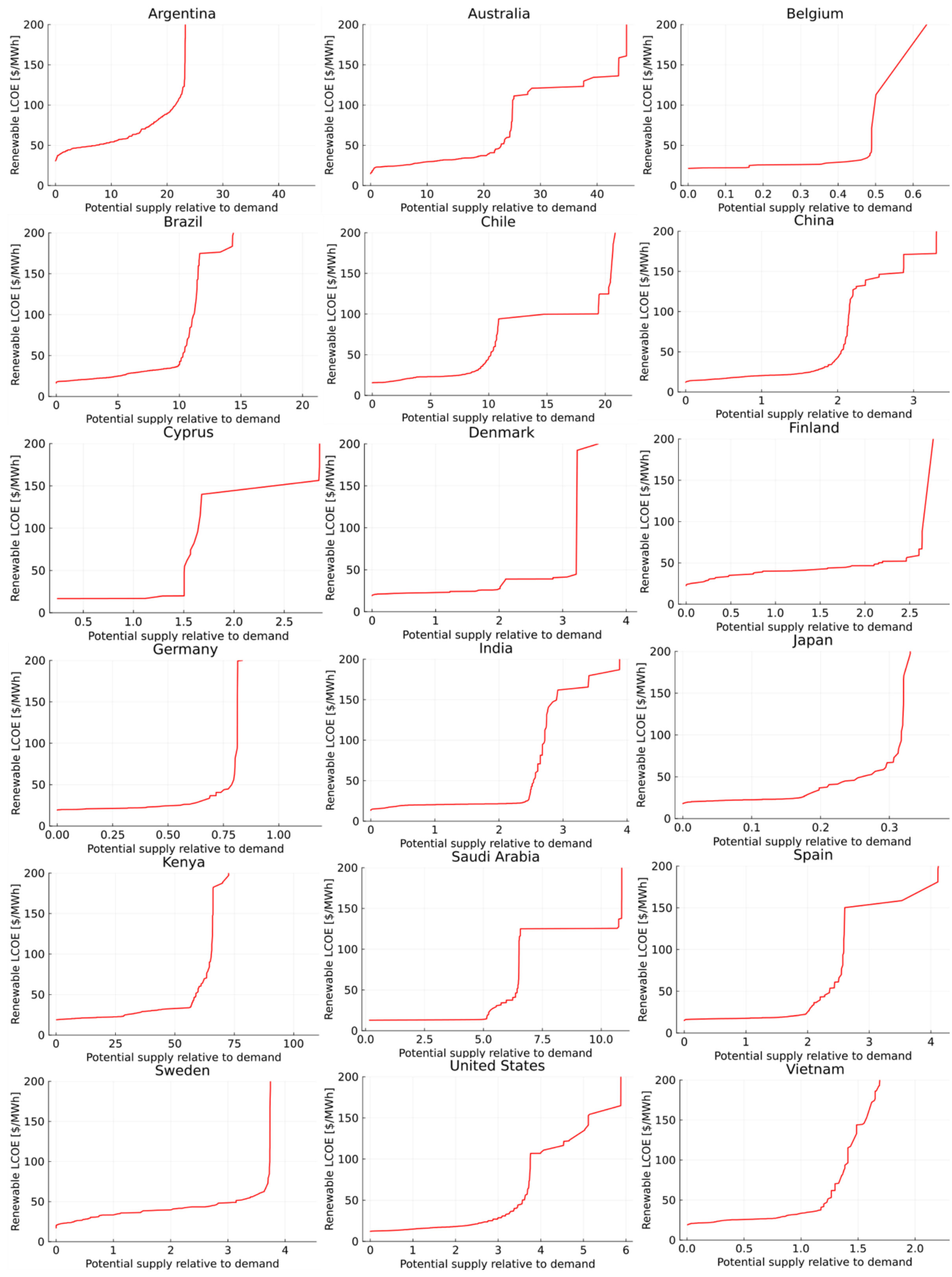

**Figure S9** National renewable energy supply curves for 18 selected countries

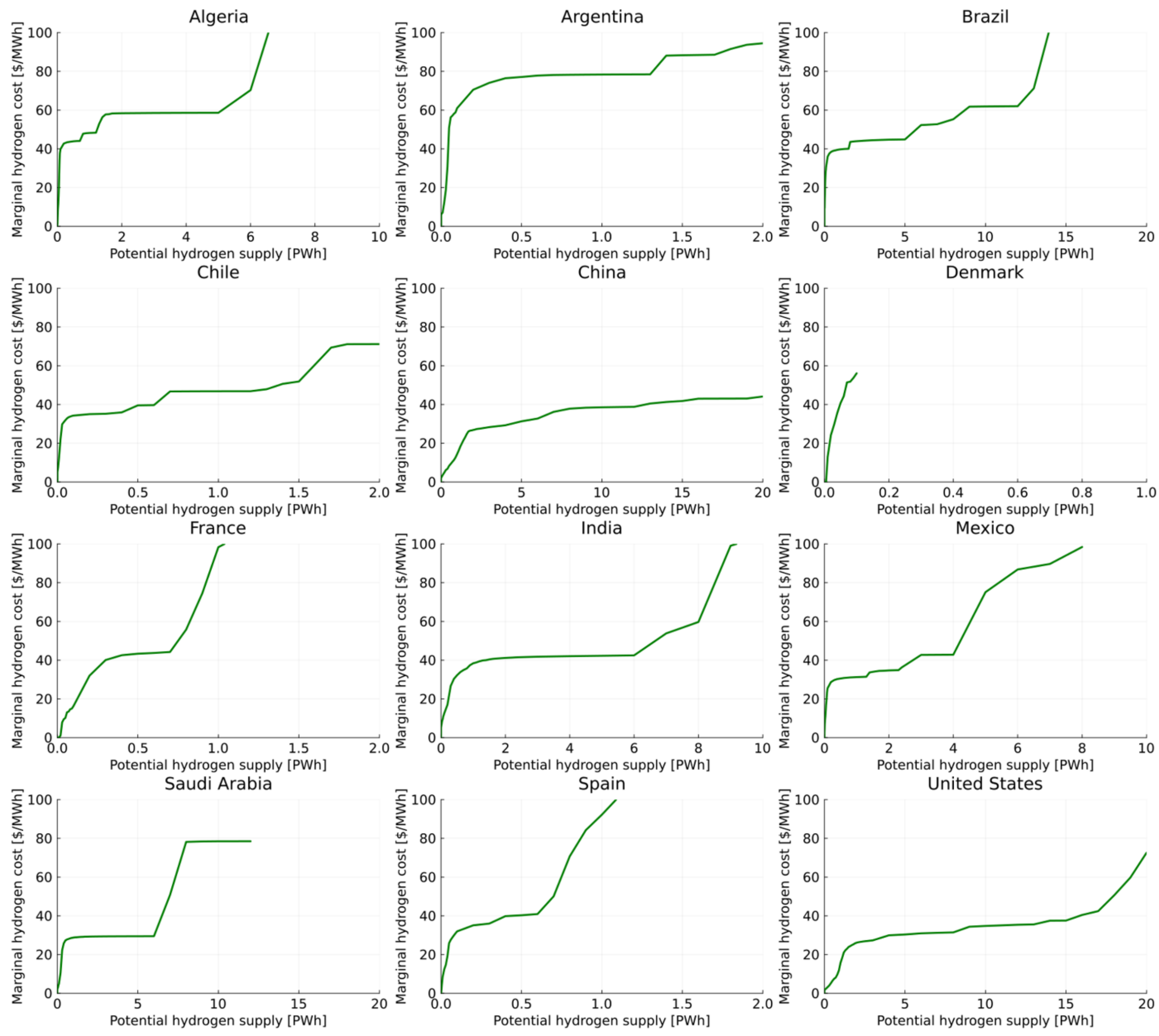

**Figure S10** National hydrogen supply curves for 12 selected countries

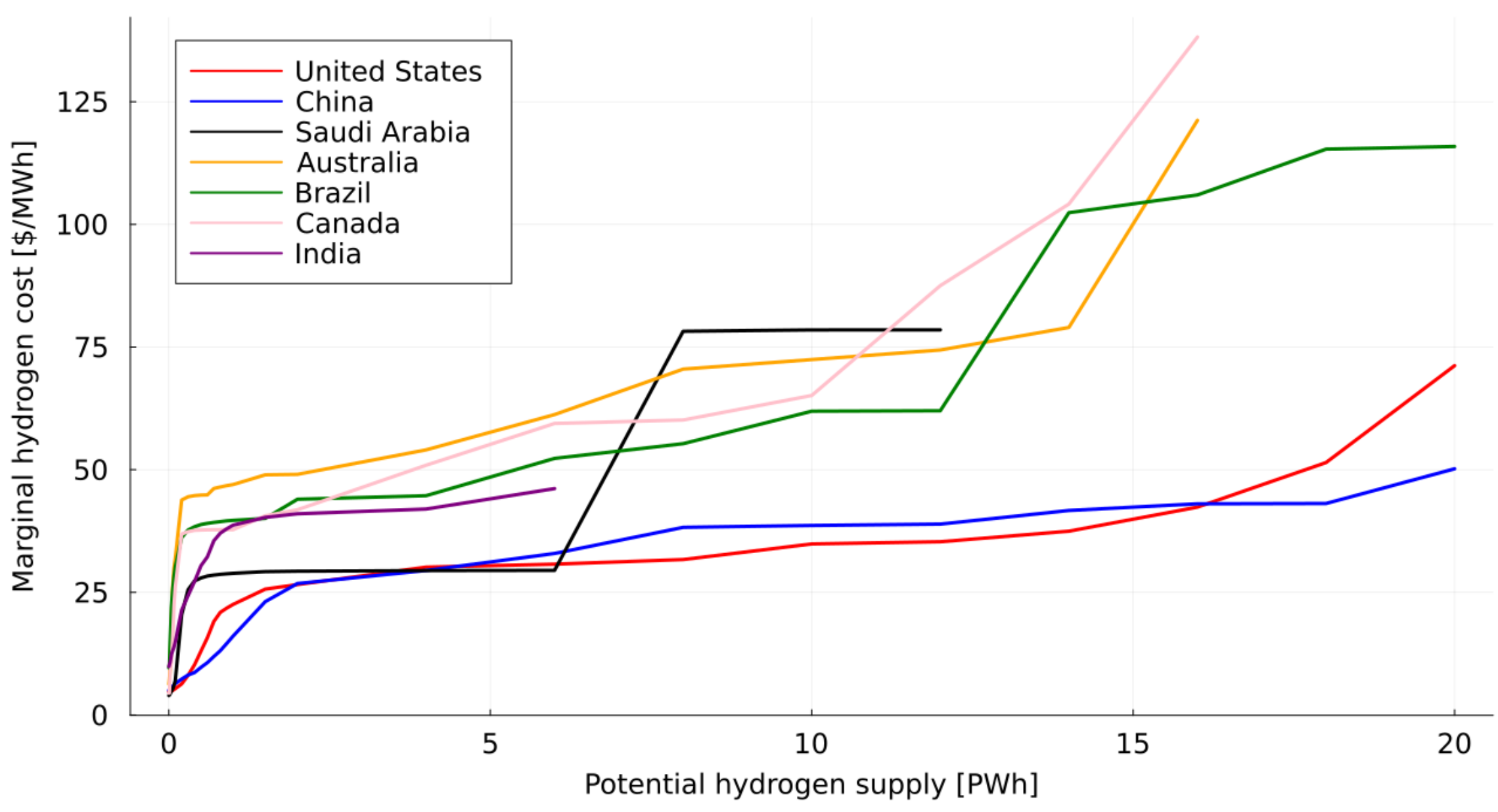

**Figure S11** National hydrogen supply curves for 7 big countries

# 2. Supplementary discussions

## 2.1 Discussion about cost assumptions and energy demand

The calculations of this study are based on the cost projections for wind and solar in 2050 by Refs. [1, 2]. We do not have endogenous learning in the model. We do see that costs are a function of cumulative production, and we further recognize that there are both local and global components of technology learning. Any simplification of the dynamics between learning rates and cost will misrepresent some aspects of the system complexity. We also recognize that electricity demand is a function of prices and that a price-mediated change in demand can have broader energy system and macroeconomic effects. However, in order to focus the attention on the impact of regional heterogeneity on a country's renewable energy self-sufficiency and export, we have chosen not to represent some known sources of global uncertainty, even though they could also affect global energy trade. For example, we assume that the cost declines implied in the IRENA estimates for 2050 adequately capture the combined effects of local and global learning for a level of deployment sufficient to achieve a 100% renewables-based energy system; we also assume inelastic demand. Explicitly representing the feedback loops between deployment, learning, cost, price, and demand requires the use of an integrated energy system model. While doing so might yield a different global equilibrium for cost and consumption, that solution would remain highly uncertain; moreover, a different global cost and consumption equilibrium might shed little additional light on the impacts of regional heterogeneity on global energy trade. Finally, we do realize that the projections of demands in 2050 are uncertain, and it is conceivable that energy-intensive end uses such as steel production and Bitcoin mining might be physically relocated to regions of low renewable energy cost, mitigating somewhat the potentials for export shown here.

## 2.2 Discussion about individual project's financial costs

It's important to note that the individual project's financial costs may vary significantly from the country-specific discount rates utilized in this study. The financial costs of a particular project include equity and debt financing expenses, along with government support such as subsidies. These components can vary widely from one project to another, across different companies and industries. However, the country-specific discount rates still provide information about relative costs when comparing one country to another, irrespective of the investor involved. For instance, a German company investing in a project in Ethiopia may secure a lower discount rate from the capital markets compared to a Tanzanian company investing in an equivalent project in Ethiopia. However, these two discount rates for investments in Ethiopia might still be higher than the corresponding rates these same two companies could obtain for another pair of equivalent projects in Italy. While any of these identical projects might secure funding, the investments in Italy would be relatively more economical than those in Ethiopia and are therefore more likely to be realized. Government investors, on the other hand, may self-finance their strategic investments, allowing them to secure effective discount rates below those available through capital markets. This is an example of a market distortion that could significantly alter renewables deployments from the patterns implied by the economic potential calculated in this study. In this study we try to illustrate the impacts of heterogenous discount rates on the location choice of the ensemble of renewable investments, rather than to evaluate returns for any specific investor or project.

## 2.3 Discussion about the methodology

Both the $RLCOE_{Ex}$ and $PEEV$ metrics are calculated based on the national renewable energy supply curve, which is derived by arranging renewable energy from low to high cost. Essentially, the $PEEV$ metric approximates the relative quantities of economically tradable renewable energy in different countries in a renewable future. This methodology aligns with the practices in competitive markets, where cost plays a key role in influencing technology deployment. However, we also recognize that various market distortions, including policies, can exert significant influence on technology deployment, potentially becoming predominant factors. In this analysis, our aim is not to predict the emergence or evolution of these distortions but rather to illustrate the economic baseline upon which any distortions would inevitably be layered. This approach doesn't necessarily forecast the future trajectory of renewable deployment, but we believe it serves a valuable policy purpose by characterizing the magnitude of policy intervention required to counteract the baseline economics of renewables deployment.

In this study, we explore a future where countries depend on domestic renewable energy resources for self-sufficiency. Self-sufficiency serves as a practical quantitative proxy for the abstract concept of energy security. We draw upon the historical and present role of energy supply security in fossil-fuel-dominated economies, where the relative balance of energy trade is a key quantitative indicator. We presume that similar energy security concerns could drive nations to seek self-sufficiency in meeting their future energy demands from domestic sources. This notion could form the basis for countries to introduce policies, altering the baseline economics illustrated in this analysis. It might also prompt countries to prioritize other zero- and negative-emission technologies, such as nuclear or (bio-) CCS. Furthermore, it could potentially reshape geopolitical alliances as countries identify new key energy trading partners. While the specific strategies individual countries might adopt to secure their future energy supplies are beyond the scope of this paper, the current energy security situation in Europe highlights the general importance of self-sufficiency in energy supply.

## 2.4 Discussion about the format of future energy trade

In this study, we intentionally maintain a neutral stance regarding the nature of future energy trade, seeking only to highlight the potential for spatial arbitrage, subject to the costs of closing the arbitrage (i.e., trade). Future energy trade could take the form of electricity directly (i.e., transmission grids) or an electricity-derived energy carrier (i.e., electro-fuel), both of which have different capital and operating costs and also different trading network topologies. The realization of any future trade in electricity or electro-fuel depends on both the potential for energy arbitrage and the cost of trade. The difference in energy costs between different countries creates incentives and opportunities for international energy trade, though we cannot predict what the form of the trade might look like. Therefore, we prioritize the analysis of the energy costs for different countries using both the simple $RLCOE_{Ex}$ metric and a comprehensive energy system model. By doing so, we aim to emphasize the fundamental economic force that will underpin future energy trade. We do not explicitly explore the trade patterns between countries and the corresponding trade benefits, but our analysis provides the basis for doing so.